\documentclass[a4paper,11pt]{article}
\usepackage{jheppub} 
\usepackage{lineno}
\usepackage{graphicx}
\usepackage{amssymb}
\usepackage{amsmath}
\usepackage{slashed}
\usepackage{dsfont}
\usepackage{tikz}
\usepackage{simplewick}
\usetikzlibrary{decorations.markings,decorations.pathmorphing}
\usepackage{longtable}
\usepackage{multirow}
\usepackage[export]{adjustbox}
\usepackage{subcaption}
\usepackage{pdflscape}
\usepackage{listings}
\usepackage{setspace}
\title{\boldmath Resonant Scattering of Boosted Dark Matter}

\author{Joshua Berger,}
\emailAdd{Joshua.Berger@colostate.edu}
\author{Zachary Orr}
\emailAdd{Zach.Orr@colostate.edu}
\affiliation{Department of Physics, Colorado State University, Fort Collins, Colorado 80523, USA}

\abstract{We develop a simulation within GENIE of the excitation of baryonic resonances by boosted dark matter. This work completes the simulation of all scattering modes for dark matter entering a detector at relativistic speeds. At some boosts, resonant scattering can contribute over 30\% to the scattering rate.  This channel offers a potentially powerful probe of the isospin structure of dark matter interactions via the relative prominence of the isospin-changing $\Delta$ resonance.  We study the estimated sensitivity of large volume detectors such as DUNE, Hyper-Kamiokande, and JUNO to all dark matter scattering modes and demonstrate the expected improvement in sensitivity when resonant scattering is included.}

\begin{document}
\tikzset{ 
		scalar/.style={dashed},
		scalar-ch/.style={dashed,postaction={decorate},decoration={markings,mark=at
				position .55 with {\arrow{>}}}},
		fermion/.style={postaction={decorate}, decoration={markings,mark=at
				position .55 with {\arrow{>}}}},
		gauge/.style={decorate, decoration={snake,segment length=0.2cm}},
		gauge-na/.style={decorate, decoration={coil,amplitude=4pt, segment
				length=5pt}}
	}
\maketitle
\flushbottom

\section{Introduction}
\label{sec:intro}

The existence of dark matter is corroborated by numerous pieces of cosmological and astrophysical evidence ~\cite{Clowe:2006eq,profumo2019introductionparticledarkmatter}, which has led in recent years to an increasingly broad set of searches for its non-gravitational interactions.  In particular, there has been a renewed emphasis of the possibility of dark matter that follows the usual Weakly Interacting Massive Particle (WIMP) story.  While direct detection of non-relativistic scattering of dark matter off nuclei remains one of the most promising avenues to discovery~\cite{LZ:2024zvo,PandaX:2024qfu,XENON:2025vwd}, there are simple, well-motivated models in which the dominant signal comes from a flux of dark matter entering terrestrial detectors at relativistic speeds, known as boosted dark matter~\cite{Agashe_2014,Berger:2014sqa,Kong:2014mia,Bringmann:2018cvk,Acevedo:2024wmx,bhalla2025supernovaboosteddarkmatterlargevolume}.  In this regime, large-volume neturino experiments such as DUNE~\cite{DUNE:2020ypp}, Super-Kamiokande~\cite{Super-Kamiokande:2002weg}, Hyper-Kamiokande~\cite{Hyper-Kamiokande:2018ofw}, JUNO~\cite{JUNO:2021vlw}, and IceCube/DeepCore~\cite{IceCube:2016zyt,IceCube-Gen2:2020qha,IceCube:2011ucd} can play a leading role.  In order to study the interactions of boosted dark matter in these experiments, a robust simulation of dark matter interactions is required.

In prior work~\cite{Berger:2018urf}, one of the authors presented an early version of a code for simulating such interactions within the standard neutrino Monte Carlo generator \lstinline{GENIE}~\cite{Andreopoulos:2009rq,andreopoulos2015genieneutrinomontecarlo}.  This code simulated scattering of boosted dark matter off of both electrons and nucleons at recoil energies above 10s of MeV, the regime relevant to most large-volume neutrino detectors.  The nucleon scattering was, however, incomplete.  It included elastic scattering, $\chi + N \to \chi + N$, and deep inelastic scattering, $\chi + N \to \chi + X(W \gtrsim 2~{\rm GeV})$, but did not include resonant scattering, $\chi + N \to \chi + N^*/\Delta$, which can dominate at intermediate energies.  Here, $\chi$ denotes the dark matter, $N$ a nucleon, $X(W \gtrsim 2~{\rm GeV})$ an arbitrary hadronic state at invariant mass larger than around $2~{\rm GeV}$, and $N^*/\Delta$ isospin $1/2$ and $3/2$ excited baryonic states.

Notably, neglecting resonant scattering is a very good approximation should the dark matter interactions with quarks be isospin conserving.  The most prominent baryonic resonance is the isospin $3/2$ $\Delta(1232)$ which cannot be produced without isospin breaking.  The calculations in \lstinline{GENIE} assume isospin conservation, but regardless $\Delta$ production should be very suppressed for a first-generation process such as resonant scattering here.  On the other hand, if boosted dark matter interactions violate isospin, resonant scattering via the $\Delta(1232)$ can be significant and would lead to a prominent, potentially reconstructable resonance in experiment.  This provides both a valuable, though not background free, search channel as well as a direct probe of the structure of the couplings of the boosted dark matter to up and down quarks.

In this work, we present an updated version of the boosted dark matter scattering code that includes resonant scattering.  During the completion of this work, the relevant hadronic matrix elements were presented in Ref.\ \cite{zink2025singlepionresonantproduction}.  Our work is complementary to this in several ways.  For one, it provides a check of their results, which we agree with.  We also have implemented these matrix elements in \lstinline{GENIE} in a version that will be folded into the publicly-available \lstinline{GENIE} code base.  Finally, we have performed a phenomenological study to quantify the effect of including resonant scattering.  We consider both an axially-coupled scalar dark matter coming from the Sun~\cite{Berger:2014sqa,Berger:2019ttc} and a vectorially-coupled fermionic dark matter raining down or up on the detector~\cite{Acevedo:2024wmx}.  It will be important to study other boosted dark matter models in the future \cite{bhalla2025supernovaboosteddarkmatterlargevolume}.  We have focused in particular on the maximally isospin violating case, where we examine the extent to which the $\Delta(1232)$ resonance can be reconstructed in the $p \pi$ channels.

The remainder of this paper is structured as follows.  In section \ref{sec:bdm-models}, we describe the modeling of the boosted dark matter flux and the interactions of the dark matter with quarks.  We then connect these interactions to hadronic interactions in the resonant regime in section \ref{sec:resonant-model}, presenting simplified calculations of the cross section.  Our phenomenological study for large-volume neutrino experiments is presented in \ref{sec:signals}.  We end with an outlook in section \ref{sec:discussion}.  Further details on the kinematics of resonant scattering can be found in appendix \ref{sec:resonant-kinematics}. Full details of the helicity amplitudes can be found in appendix \ref{sec:helicity-amp}.

\section{Boosted Dark Matter Models}
\label{sec:bdm-models}

In order to develop a projected sensitivity to boosted dark matter (BDM), both the BDM flux and a model of the interactions in the detector are required.  The module presented in this work develops the interactions in the detector in which the dark matter scatters resonantly off of a nucleon, giving a excited baryon resonance.  We further consider two different BDM flux models~\cite{Berger:2014sqa,Berger:2019ttc,Acevedo:2024wmx} to demonstrate the extra sensitivity gained by including this scattering channel. In this section, we describe the flux and quark interaction models developed for the phenomenological study.  While these are interesting benchmarks, it is worth noting that the code is sufficiently flexible as to accommodate both scalar and fermionic dark matter interacting with arbitrary parity and isospin structure.  At the moment, the code is however limited to scattering in the detector mediated by a vector boson.

In the first model we consider~\cite{Berger:2014sqa,Berger:2019ttc,Agashe_2014}, there are two components of dark matter, with the heavy component dominating the abundance of relic dark matter. This component gets captured by the Sun and accumulates therein, leading to an enhanced annihilation of the heavy dark matter, in the center of the Sun. The heavier dark matter component annihilates into lighter dark matter particles which receive a Lorentz boost. We will refer to this model as the two-component model. 

In the other BDM model \cite{Acevedo:2024wmx}, a long range force at roughly the scale of the Earth accelerates a single component of dark matter as it approaches the Earth. We will refer to this as a the dark matter rain model. It has been shown, using elastic and deep inelastic scattering alone, that experiments such as the DUNE far detector, Super-Kamiokande, Hyper-Kamiokande, JUNO, and IceCube should all have sensitivity to both of these flux models beyond direct detection constraints for some parameters ~\cite{Acevedo:2024wmx,Berger:2019ttc,Berger:2014sqa,DUNE:2015lol,JUNO:2015zny,Super-Kamiokande:2002weg,Hyper-Kamiokande:2018ofw}.

For the purposes of detection, we focus on scalar dark matter with an axial coupling to quarks in the case of the two-component model. This is the scenario in which the large volume neutrino experiments above have leading sensitivity \cite{Berger:2019ttc}. For the dark matter rain model, at sufficient boost, these neutrino experiments always have leading sensitivity \cite{Acevedo:2024wmx}, so for simplicity we consider fermion dark matter with a vector coupling to quarks. In both models, the coupling is mediated by a heavy vector boson that we denote by $Z^\prime$ which couples to the dark matter and SM sectors with a coupling strength of $g_{Z^{\prime}}$. For both scenarios, we consider an isospin conserving and a maximally isospin violating coupling to quarks. In the latter scenario, resonant scattering is significantly enhanced as the $\Delta$ baryons can only be produced if there is isospin violation. As our modeling of the baryons assumes isospin is a good symmetry of the hadronic sector, the only source of isospin breaking could be in the couplings of the mediator. 

Below we describe both the flux and interaction modeling for each of these two scenarios.
 
\subsection{Two Component Model}

In the two component model \cite{Berger:2014sqa,Berger:2019ttc}, both components of dark matter are taken to be scalar particles.
The Lagrangian for a scalar DM particle, $\phi_i$, and quarks, $q$, with the vector mediator, $Z^\prime$, includes an interaction term,

\begin{equation}
    \mathcal{L}_{\text{int}} = g_{Z^\prime}Z^\prime_{\mu} \sum_i Q_i (i\phi^\dagger_i \partial^\mu \phi_i + \text{h.c.}) + g_{Z^\prime} Z^\prime_\mu \sum_q Q_q \overline{q} \gamma^\mu (Q_{L,q} P_L + Q_{R,q} P_R) q,
\end{equation}
where $i = (\psi,\chi)$ labels the heavy and light component of dark matter respectively and $q = (u,d,s,c)$ are the relevant quarks.

The dominant component with a heavy mass of $m_\psi$, after having been captured by scattering off atomic nuclei in the sun, annihilates to a subdominant component with a lighter mass of $m_\chi$. The subdominant component of the DM escapes the sun with a Lorentz boost
\begin{equation}
    \gamma = \frac{m_\psi}{m_\chi},
\end{equation}
 which can result in relativistic speeds for a sufficiently large mass ratio \cite{Berger:2019ttc}. The resulting solar flux of the BDM is an ideal candidate for study at large volume neutrino detectors.

In this model, the light dark matter $\phi_\chi$ has a normalized charge of $Q_\chi = 1$. As a benchmark, the mediator mass is taken to be $m_{Z^{\prime}}= 1~\text{GeV}$. The quark charges are chosen such to be axial, with $Q_{L,q} = -Q_{R,q}$ for all quark flavors. The gauge coupling $g_{Z^\prime}$, the mass of the lighter dark matter $m_\chi$, and the boost $\gamma$ given by the ratio of the two dark matter masses are left as free parameters of the model. 

A key feature of this model is the directional dependence of the BDM flux as it passes through the detector. The signature of this model is seen in the angle, $\theta$, of the reconstructed momenta of visible particles relative to the direction from the Sun to the Earth. The quantity $\cos{\theta}$ is the primary measure used for discriminating against the nearly isotropic background from atmospheric neutrinos in this model. In an experimental search, neutrinos reconstructed as going toward the Sun may be used as a sideband to reduce systematic uncertainties.

An important constraint on this model comes from spin-dependent direct detection searches for non-relativistic DM which constrain the cross-section for capture processes, $\sigma_{DD}$, in the Sun\cite{PhysRevLett.122.141301}. In the regime of interest, where $m_\psi \gtrsim 4~\text{GeV}$ and the annihilation cross section required to obtain the observed relic abundance of dark matter $(\sigma v)_{ann} \approx 3 \times 10^{26}~\text{cm}^3/\text{s}$, the capture rate $C$ equilibrates with the annihilation rate $A N^2$ during the lifetime of the sun, of order 1 billion years. Here, $N$ is the amount of captured dark matter. Additionally, re-scattering remains negligible giving rise to a mono-energetic flux of BDM ~\cite{Berger:2014sqa,Berger:2019ttc}. In this way, the BDM flux from the sun is directly proportional to the capture rate of the heavy dark matter. The capture rate is, in turn, directly proportional to the DM-Nucleon cross-section \cite{Berger:2014sqa} and constrained by direct detection results~\cite{PICO:2019vsc,LZ:2024zvo,PandaX:2024qfu,XENON:2025vwd}. To simulate the flux of solar BDM, the \lstinline{SolTrack} package~\cite{PhysRevD.70.023006} with an implementation in \lstinline{gsolflux}~\cite{Berger:2018urf}. 

Previous work on this model showed a LArTPC reference detector with sensitivity to $g_{Z^\prime}^4 = (1.54 - 22.0) \times 10^{-7}$ at two standard deviations over the selected parameter space~\cite{Berger:2019ttc}.

\subsection{Dark Matter Rain}

In this model there is only a single component of dark matter $\chi$, which is taken to be fermionic for concreteness, though nothing prevents one from considering scalar dark matter as well. The short-range interactions are again mediated with a $Z^\prime$ mediator,

\begin{equation}
\mathcal{L}_{\text{int}} = g_{Z^\prime}Z^\prime_{\mu}\bar{\chi}\gamma^{\mu}(Q_{L,\chi} P_{L}+Q_{R,\chi} P_{R})\chi + g_{Z^\prime} Z^\prime_\mu \sum_q Q_q \overline{q} \gamma^\mu (Q_{L,q} P_L + Q_{R,q} P_R) q.
\end{equation}

In addition to the short-range force that determines interactions in the detector, there must be a long range force that accelerates the dark matter as it approaches the earth~\cite{Acevedo:2024wmx}. The long range force must have a coupling to both the SM and $\chi$. The SM coupling is limited by fifth force searches \cite{Acevedo:2024wmx}. The $\chi$ coupling is limited, dominantly, by constraints from the bullet cluster \cite{Markevitch:2003at,Clowe:2006eq,Davoudiasl:2017pwe}. Nevertheless, one can find parameter space with a significant boost due to this force. This force may be mediated by a scalar or by a vector. However, the scalar case leads to a runaway acceleration at sufficiently large couplings. Nevertheless, at low dark matter masses, relativistic boosts can easily be achieved and the dark matter ends up essentially mono-energetic~\cite{Acevedo:2024wmx}. As a benchmark, the short range mediator is taken to have a mass of $m_{Z'}=10\ \text{TeV}$. In this model, consider a vector coupling for the dark matter as the sensitivity of large volume experiments is far better than that of dedicated direct detection experiments over the entire parameter space of interest.  For all quark flavors, we take $Q_{L,q} = Q_{R,q}$.  As in the two-component model, the gauge coupling $g_{Z^\prime}$, the mass of the lighter dark matter $m_\chi$, and the boost $\gamma$ are left as free parameters.

As is the case for the two-component model, the background here is dominated by the nearly isotropic atmospheric neutrino flux. In this case, the dark matter is expected to pass through the detector along an essentially vertical path (either down or up), hence the name dark matter rain. Since the boost is taken to be relativistic, the initial velocity of the dark matter is subdominant in the ultimate kinematics, which are dominated by the vertical long range force at the surface of the earth \cite{Acevedo:2024wmx}.  For this reason, the angle, $\theta$, of the reconstructed visible particles is defined to be with respect to the vertical axis for this model. To simulate an appropriate flux, the incident DM particle beam angle has been rotated into the vertical axis of the detector.  

The parameter space previously explored by the Dark Matter Rain model reported a sensitivity to the BDM cross sections of $10^{-53}$ to $10^{-60} \text{cm}^2$ for sub-GeV dark matter and boosts as high as 1000 \cite{Acevedo:2024wmx}.

\section{Resonant Scattering Model}
\label{sec:resonant-model}

The main focus of this section is to determine the partial cross-section, for the two models described above, undergoing a resonant nuclear scattering process by which the incident dark matter particle is exciting the nucleon target to a higher mass baryonic state. Furthermore, these calculations are utilized to implement a simulation of this process in \lstinline{GENIE}~\cite{Andreopoulos:2009rq,andreopoulos2015genieneutrinomontecarlo}.  

The scattering amplitude from interactions between dark matter particles and nucleons in the intermediate final state invariant mass range of $W = 1.2-2~\text{GeV}$ can have a significant contribution from processes producing excited baryon states. This is relevant when dark matter is sufficiently energetic to produce an invariant mass of that order, in other words 
\begin{equation}
    \gamma \gtrsim 1.3 + \frac{0.3~\text{GeV}}{m_\chi}
\end{equation}

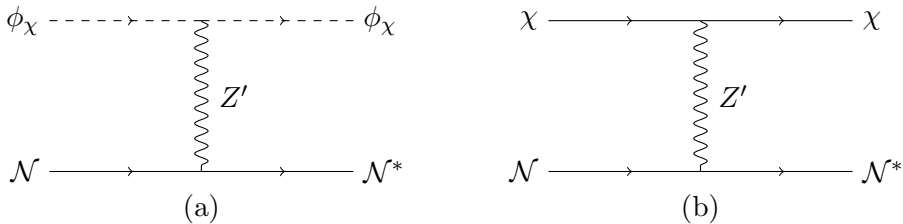
\begin{figure}[!tbh]
\begin{center}
    \begin{tikzpicture}
\draw[scalar-ch] (0,0) node[left] {$\phi_\chi$} -- ++ (-0:2cm) coordinate (v1);
\draw[scalar-ch] (v1) -- ++ (180:-2cm) node[right] {$\phi_\chi$};
\draw[gauge] (v1) -- node[right=0.1cm] {$Z^\prime$} ++ (270:2cm) coordinate (v2);
\draw[fermion] (v2) -- ++ (0:2cm) node[right] {$\mathcal{N}^*$};
\draw[fermion] (v2) ++ (0:-2cm) node[left] {$\mathcal{N}$} -- ++ (0:2cm);
\node at (2,-2.5) {(a)};
\end{tikzpicture}\hspace{1cm}
    \begin{tikzpicture}
\draw[fermion] (0,0) node[left] {$\chi$} -- ++ (-0:2cm) coordinate (v1);
\draw[fermion] (v1) -- ++ (180:-2cm) node[right] {$\chi$};
\draw[gauge] (v1) -- node[right=0.1cm] {$Z^\prime$} ++ (270:2cm) coordinate (v2);
\draw[fermion] (v2) -- ++ (0:2cm) node[right] {$\mathcal{N}^*$};
\draw[fermion] (v2) ++ (0:-2cm) node[left] {$\mathcal{N}$} -- ++ (0:2cm);
\node at (2,-2.5) {(b)};
\end{tikzpicture}

\end{center}
\caption{The Feynman diagram for (a) scalar dark matter and (b) fermionic dark matter undergoing resonant scattering off a nucleon $\mathcal{N}$.  A baryonic excitation $\mathcal{N}^*$ is produced. Where, $Z^\prime$ is the gauge boson that mediates the interaction.}\label{fig:feynman}
\end{figure}

The Feynman diagram for fermionic and scalar dark matter resonant scattering off of a nucleon is shown in Fig.~\ref{fig:feynman}.  In this process, a nucleon $\mathcal{N}$ is up-scattered into a baryonic resonance $\mathcal{N}^*$.  The amplitude for such a process can be written in the form
\begin{equation}
    i\mathcal{M} \propto \left(-g_{\mu\nu} + \frac{q_\mu q_\nu}{m_{Z^\prime}^2}\right) J^\mu_l J^\nu_{h},
\end{equation}
where $J_l$ and $J_h$ are the ``leptonic'' and hadronic currents while $q$ is the momentum carried by the virtual $Z^\prime$ boson. Despite its name, the leptonic current characterizes the structure of dark matter interactions with the mediator $Z^\prime$.

For an interaction mediated by an intermediate vector boson of virtual mass $q^2$, it is convenient to decompose these currents along the polarization basis vectors $e_\mu$ \cite{Ravndal:1973xx}. In the massless limit, $e^\mu q_\mu = 0$ and the polarization basis can be expressed in a frame where the momentum transfer is along the z-axis as $q = (\nu;0,0,Q)$ ~\cite{Ravndal:1973xx,Rein:1980wg},
\begin{eqnarray}
    e_S^\mu &=& \sqrt{\frac{1}{-q^2}}(Q;0,0,\nu)\\
    e_R^\mu &=& \sqrt{\frac{1}{2}}(0;-1,-i,0)\\
    e_L^\mu &=& \sqrt{\frac{1}{2}}(0;+1,-i,0).
\end{eqnarray}
For a leptonic current which includes an axial component, the axial-vector divergence is non-zero and proportional to the mass of the dark matter.  In this case, the massless particle approximation is invalid and a fourth basis vector must be included for which $e^\mu q_\mu \neq 0$.  To keep the normalization scheme adopted by previous work ~\cite{Ravndal:1973xx,Rein:1980wg}, we take this vector to be 
\begin{equation}
    e_q^\mu = -\sqrt{\frac{1}{-q^2}}q^\mu.
\end{equation}
Note that alternative approaches to massive fermions are possible ~\cite{zink2025singlepionresonantproduction} in which polarization-dependent currents are constructed and the current for each polarization is decomposed in terms of non-trivial components.

The resulting partial cross section for this $2\rightarrow 2$ scattering may be written in the general form,
\begin{equation}
\frac{d^2 \sigma}{d q^2 d W} = \sigma_0 [A_L \sigma_L + A_R \sigma_R + A_S \sigma_S + A_q \sigma_q],
\end{equation}
where $q^\mu$ is the momentum transfer four-vector described above and $W$ is the invariant mass of the final state baryonic resonance. The requirement of on-shell resonance production can be easily relaxed by including a Breit-Wigner factor, so we keep the expression general. 
 The full derivation of this expression is given in \ref{sec:resonant-kinematics}.

The kinematic prefactor of this expression is given by,
\begin{equation}
\sigma_0 = \frac{g_{Z^\prime}^4}{16 \pi (1 - m_\chi^2/E^2) (q^2 - m_{Z'}^2)^2} \frac{W}{m_\mathcal{N}} \frac{-q^2}{Q^2},
\end{equation}
where $E$ is the energy of the incident dark matter in the lab frame. The mass of the target nucleon, the incident dark matter and the intermediate vector boson are written as $m_\mathcal{N}$, $m_\chi$ and $m_{Z^\prime}$, respectively. 

The coefficients $A_i$ of the cross section structure factors $\sigma_i$ depend on the type of dark matter considered. For fermion DM,
\begin{subequations}
    \begin{equation}
        A_L = (Q_L^\chi)^2 u^2 
+ (Q_R^\chi)^2 v^2
+\frac{2m_\chi^2 Q^2 Q_L^\chi Q_R^\chi}{ E^2 q^2};
    \end{equation}
    \begin{equation}
        A_R = (Q_L^\chi)^2 v^2 
+(Q_R^\chi)^2 u^2
+\frac{2m_\chi^2 Q^2 Q_L^\chi Q_R^\chi}{ E^2 q^2};
    \end{equation}
    \begin{equation}
        A_S =  2 u v ((Q_L^\chi)^2 + (Q_R^\chi)^2)
+\frac{m_\chi^2 Q^2 (Q_L^\chi - Q_R^\chi)^2}{ E^2 q^2};
    \end{equation}
    \begin{equation}
        A_q = - \frac{m_\chi^2 Q^2 (Q_L^\chi - Q_R^\chi)^2 (m_{Z^\prime}^2 - q^2)^2 }{E^2 m_{Z^\prime}^4 q^2}.
    \end{equation}
\end{subequations}

On the other hand, for scalar DM,
\begin{subequations}
    \begin{equation}
        A_L = A_R = (Q_S^\chi)^2\left( 2 u v + \frac{2m_\chi^2 Q^2}{q^2 E^2} \right);
    \end{equation}
    \begin{equation}
        A_S = (Q_S^\chi)^2(u + v)^2;
    \end{equation}
    \begin{equation}
        A_q = 0.
    \end{equation}
\end{subequations}

The kinematic variables $u$ and $v$ are given by
\begin{subequations}
    \begin{equation}
        u = \frac{E + E^\prime + Q}{2 E^\prime},
    \end{equation}
    \begin{equation}
        v = \frac{E + E^\prime - Q}{2 E^\prime},
    \end{equation}
\end{subequations}
where $E^\prime$ is the energy of the outgoing DM in the lab frame.

The structure factors $\sigma_i$ can be expressed directly in terms of the helicity amplitudes for resonance production. These factors are the independent of the nature of the dark matter and are given by
\begin{subequations}
    \begin{equation}
         \sigma_{L}=\frac{M}{m_{\mathcal{N}}}\frac{1}{2}(|f_{+3}|^2 + |f_{+1}|^2) \delta (W - M);
    \end{equation}
    \begin{equation}
        \sigma_{R}=\frac{M}{m_{\mathcal{N}}}\frac{1}{2}(|f_{-3}|^2 + |f_{-1}|^2) \delta (W - M);
    \end{equation}
    \begin{equation}
        \sigma_{S}=\frac{m_{\mathcal{N}}}{M}\frac{Q^{2}}{-q^2}\frac{1}{2}(|f_{0+}|^2 + |f_{0-}|^2) \delta (W - M);
    \end{equation}
    \begin{equation}
        \sigma_{q}=\frac{m_{\mathcal{N}}}{M}\frac{Q^{2}}{-q^2}\frac{1}{2}(|f_{D+}|^2 + |f_{D-}|^2) \delta (W - M),
    \end{equation}
\end{subequations}
where $M$ is the resonance mass and the amplitudes $|f|^2$ denote helicity amplitudes that are described further below. The delta function may be replaced by a Breit-Wigner factor in the narrow width approximation.

The helicity amplitudes $f_{\pm 2j}$ and $f_{0\pm}$ may be calculated in the same way for the case of neutrino scattering~\cite{Ravndal:1973xx,Rein:1980wg}. The additional structure factor, $\sigma_q$, arises from the inclusion of massive dark matter with an axial coupling. Following \cite{Ravndal:1973xx,Rein:1980wg}, the new helicity amplitudes may be expressed as:
\begin{equation}
f_{D\pm}= \left\langle \mathcal{N}, \pm \frac{1}{2} \right| F_{D\pm} \left| \mathcal{N}^*, \pm \frac{1}{2} \right\rangle.
\end{equation}
Where $F_D$ is the divergent helicity operator associated with the divergent component of the resonance production current operator. The divergence is defined from the PCAC theorem \cite{Ravndal:1973xx}, expressed as
\begin{equation}
    q^\mu F_\mu = q^\mu F_\mu^V - q^\mu F_\mu^A = 0 - D,
\end{equation}
where $D$ is the divergence.

The approach for neutrino scattering, based on the model by Feynman, Kislinger, and Ravndal~\cite{Feynman:1971wr}, describes the baryons by a relativistic harmonic oscillator potential. This model is then used to calculate the transition amplitudes to excited baryonic states. The general hermitian operator for such an excitation may be expressed in terms of a vector and axial-vector component. Following the conventions of prior work,~\cite{Ravndal:1973xx,Rein:1980wg},
\begin{equation}
    F_\mu = F_\mu^V - F_\mu^A = \frac{1}{2M}(V_\mu- A_\mu),
\end{equation}
where the vector and axial-vector currents, $V_\mu$ and $A_\mu$, contain quark charges, $e_a$ and $e_a^\prime$, respectively. The projection of these vector and axial-vector currents along a basis of polarization vectors is derived by \cite{Feynman:1971wr} and tabulated as helicity amplitudes for neutrino interactions by \cite{Rein:1980wg,Ravndal:1973xx}.
The helicity operator of interest, $F_D$, can be determined in the resonance rest frame in terms of the polarization basis projection on the axial-vector current as ~\cite{Ravndal:1973xx,Feynman:1971wr},
\begin{equation}
    F_D = \frac{Q^*}{Q^{*2}}\frac{q^\mu A_\mu}{2M}
\end{equation}
where, in the resonance rest frame, $q^\mu = (\nu^*;0,0,Q^*)$ and the axial-vector current, normalized by an undetermined constant $Z$,
\begin{equation}
    A_\mu = 9 Z e'_a (\slashed p_a \gamma_\mu \gamma_5 e^{iq\cdot u_a} + \gamma_\mu \gamma_5 e^{iq\cdot u_a} \slashed p_a)
\end{equation}
with momentum and coordinate operators, $p_a$ and $u_a$ \cite{Feynman:1971wr}, acting on quark $a$ of the triplet in the resonance rest frame. For resonant production of excitations from an initial ground-state baryon, the divergent helicity operator can be written in the convenient and conventional form,
\begin{equation}
   F_D = -9 e'_a \left[ C_D \sigma_{az} + B_D (\vec{\sigma_a}\cdot \vec{a}) \right]e^{-\lambda a_z}
\end{equation}
with,
\begin{eqnarray}
    C_D &=& \frac{ZG_A}{3}\left(1 + \frac{3q^2 + N\Omega}{(M+m_\mathcal{N})^2 - q^2}\right)\\
    B_D &=& \frac{ZG_A}{2MQ^*}\sqrt{\frac{\Omega}{2}}\left(M - m_\mathcal{N} \right)
\end{eqnarray}
where $N$ is the number of excitations in $\mathcal{N}^*$ and $\Omega = 1.05\ [\text{GeV}]^2$ is the harmonic oscillator constant \cite{Ravndal:1973xx}. The transition form factor $G_A$ is given in \ref{sec:resonant-kinematics}. The components of the excitation matrix are then calculated in the resonance rest frame by the constituent quark model dynamics \cite{Feynman:1971wr}.

The matrix elements are determined by products of operators acting on isospin, $e_a$, ordinary spin, $\sigma_a$, and orbital motion, $a$. The operators for spin and orbital motion are known \cite{Feynman:1971wr} and will not be reproduced here. The quark charges are model dependent, so the $SU(2)$ operator for non-strange resonances takes the form,
\begin{eqnarray}
    e_a &=& \frac{Q_u^V - Q_d^V}{2}\tau_3 + \frac{Q_u^V + Q_d^V}{2}\mathds{1},\\
    e'_a &=& \frac{Q_u^A - Q_d^A}{2}\tau_3 + \frac{Q_u^A + Q_d^A}{2}\mathds{1},
\end{eqnarray}
where the vector and axial-vector charges are related to the left- and right-handed charges by, 
\begin{eqnarray}
    Q_f^V &=& \frac{Q^f_{R} + Q^f_L}{2},\\
    Q_f^A &=& \frac{Q^f_R - Q^f_L}{2}.
\end{eqnarray}

Further details of the calculation can be found in Appendix \ref{sec:resonant-kinematics}.  The full helicity amplitudes for the hadronic currents are tabulated in Appendix \ref{sec:helicity-amp}.

\begin{figure}[!tbh]
\centering
        \begin{subfigure}{0.49\textwidth}
        \includegraphics[scale=0.48]{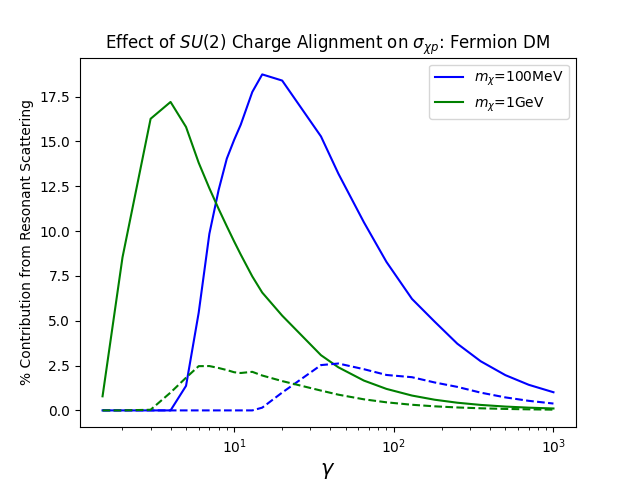}
       \label{fig:RES_Ferm}
       \end{subfigure}
       \begin{subfigure}{0.49\textwidth}
       \includegraphics[scale=0.48]{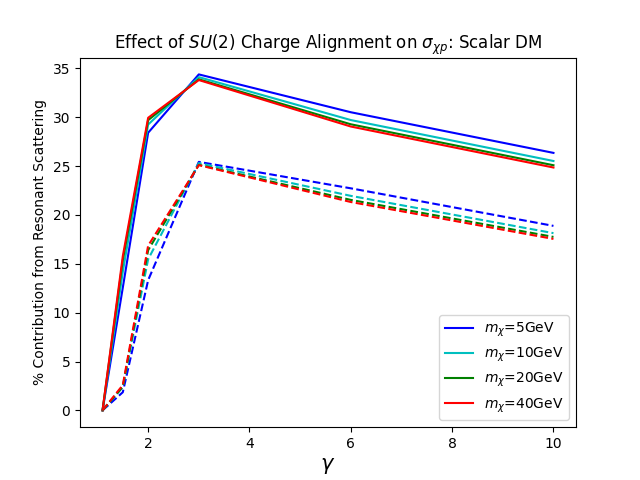}
       \label{fig:RES_Scal}
        \end{subfigure}
\caption{The fractional effect of $SU(2)$ charge alignment to the contribution of resonant scattering in the proton cross-section of $\sigma_{\chi p}$ in terms of the boost factor $\gamma$. Solid lines represent the maximal isospin symmetry breaking model while dashed lines are the isospin symmetry conserving model. The left plot shows the two mass points considered in the DM Rain model, while the right plot represents the various mass points taken for the two-component DM model.}
\label{Iso_Res_cont}
\end{figure}

The relative isospin charges have a significant effect on the resonant scattering contribution \ref{Iso_Res_cont}. The helicity amplitudes of the hadronic current depend on isospin breaking operators which acts on the ground state of the target nuclei. From the helicity amplitude tables \ref{sec:Helicity_table}, it can be shown that the largest contribution to the resonant signal is from the $\Delta$ baryon resonance. The helicity amplitudes for this transition are suppressed when $u$ and $d$ quark charges are aligned in the helicity basis. 

\begin{table}[!tbh]
\centering
\setlength{\tabcolsep}{12pt}
        \begin{tabular}{c|c c |c c}
            \hline
                Chiral Structure & \multicolumn{2}{c|}{Vector} & \multicolumn{2}{c}{Axial} \\
                \hline
              Isospin Structure & Min & Max & Min & Max \\
              \hline
               $(Q_L^{(u)},Q_R^{(u)})$ & (1,1) & (1,1) & (1,-1) & (1,-1)\\
              $(Q_L^{(d)},Q_R^{(d)})$ & (1,1) & (-1,-1) & (1,-1) & (-1,1) \\
              $(Q_L^{(s)},Q_R^{(s)})$ & (1,1) & (-1,-1) & (1,-1) & (-1,1) \\
              $(Q_L^{(c)},Q_R^{(c)})$ & (1,1) & (1,1) & (1,-1) & (1,-1)\\
        \hline
        \end{tabular}
\setlength{\tabcolsep}{6pt}
        \label{tab:charges}
        \caption{Table of left- and right-handed quark charges for isospin conserving (Min) and maximal isospin symmetry breaking (Max) in the two BDM models considered, where quark couplings are either vector (left two columns) or axial (right two columns).}
\end{table}

In Fig.~\ref{Iso_Res_cont}, the minimal and maximal isospin symmetry conserving currents are simulated to demonstrate the relative contribution to the scattering amplitude. To compare the reach of various experimental searches below, a model with maximal isospin symmetry breaking is used to demonstrate the largest possible effect of resonant scattering.  We consider such potential searches in the next section.

The model described in this section has been implemented into the \lstinline{GENIE} neutrino Monte Carlo code and will be merged into a future release version of \lstinline{GENIE}~\cite{andreopoulos2015genieneutrinomontecarlo}.  Provided \lstinline{GENIE} is built with the \lstinline{--enable-boosted-dark-matter} option, resonant boosted dark matter scattering is included by default.  A detailed description of the standard options for running the code can be found in \cite{Berger:2018urf}.  One new feature is that we offer the event generator list \lstinline{NonRes} to omit resonant scattering.  The default DM event generator list, \lstinline{DM}, includes resonant scattering.

\section{Signals in Large Volume Experiments}
\label{sec:signals}

Large volume neutrino experiments such as DUNE, Super-Kamiokande, Hyper-Kamiokande, JUNO, and IceCube/DeepCore are expected to have the leading sensitivity to boosted dark matter models at sufficiently high boost.  In this section, we perform the first phenomenological study of boosted dark matter at these experiments including resonant scattering. For this study, we have generated both a set of signals and of atmospheric neutrino backgrounds using the same nuclear model \lstinline{GDM18_00a_00_000} in \lstinline{GENIE}. The signal is genrerated using  \lstinline{GDM18_00a_00_000} for fermionic DM and \lstinline{GDM18_00b_00_000} for scalar DM, though these models differ only in the spin of the DM.

The only detectors chosen for this study are those for which the intermediate scattering plays a significant role in contributing to the signal over the range of parameters chosen in these models. For detectors such as IceCube and DeepCore the threshold for detection is too high for resonant scattering to be of interest. Detector thresholds for LZ are much lower than the particles produced by these interactions and so the intermediate scattering regime is subdominant.

We generate 10,000 signal events at various boosts and masses.  The background signal is broken into low energy and high energy samples to ensure large enough statistics at high energies. Though there are fewer high energy neutrinos, these neutrinos are more likely to pass selection cuts targeted at boosted dark matter.  The atmospheric neutrino flux model we use is the HAKKM 2014 model~\cite{Honda:2015fha}. The combined background sample is formed by 500,000 events from atmospheric neutrinos with the energies $E_\nu < 10 \text{ GeV}$ and 500,000 events within the range of $10 < E_\nu < 1000 \text{ GeV}$.

\begin{table}[!tbh]
\centering
        \begin{tabular}{c|c c c c c}
        \hline
            Experiment & $\mu^\pm$ (MeV) & $\pi^\pm$ (MeV) & $p$ (MeV) & $e^\pm$ (MeV) & $\gamma$ (MeV) \\
            \hline
              DUNE & 35 & 35 & 80 & 30 & 30  \\
              Super-/Hyper-K  & 55 & 75 & 485 & 3 & 3 \\
              JUNO & 0.5 & 0.5 & 0.5 & 0.5 & 0.5  \\
        \hline
        \end{tabular}
        \caption{Thresholds used for phenomenological sensitivities of the large volume neutrino experiments to BDM.}        \label{tab:thresholds}
\end{table}
The signal is reconstructed from the four-momentum of visible final state particles, summed together coherently. The visible particle kinetic energy thresholds are based on the detector sensitivities used in Ref. \cite{Acevedo:2024wmx}.  The thresholds used are shown in Table \ref{tab:thresholds}. For DUNE, Hyper-K and JUNO the exposure times are predicted to be 10 years while Super-K has been ongoing since 1996 so the exposure time utilized for this study is 28 years.

The only visible particles considered are protons, charged and neutral pions, muons, photons, and electrons. The selection criteria employed here do not include neutrons since they are typically challenging to reconstruct in detectors.

The background for these simulations is primarily atmospheric neutrinos ~\cite{Berger:2019ttc,Acevedo:2024wmx}, which are taken to have an approximately isotropic distribution in the detector.  Rare cosmic rays are also a possible confounding factor that may need to be considered, but such consideration is beyond the scope of this work. While the dark matter is expected to be incident from particular directions in many models, e.g.\ the Sun and the vertical in this work, the atmospheric neutrino background is nearly isotropic.  By looking at events that appear from the expected BDM direction and those that do not, a strong handle can be obtained on the background. 
 ``Wrong direction'' events can be used as a sideband to constrain the background rate in the expected direction.  Furthermore, selecting a narrow angular window around the expected dark matter direction is a powerful tool for reducing the background.  The isotropic distribution of the incident atmospheric neutrinos gives a reliable side band measurement. We make use of this directional discrimination in our study.

    \begin{figure}[!tbh]
    \centering
        \begin{subfigure}{0.45\textwidth}
        \includegraphics[width=\linewidth]{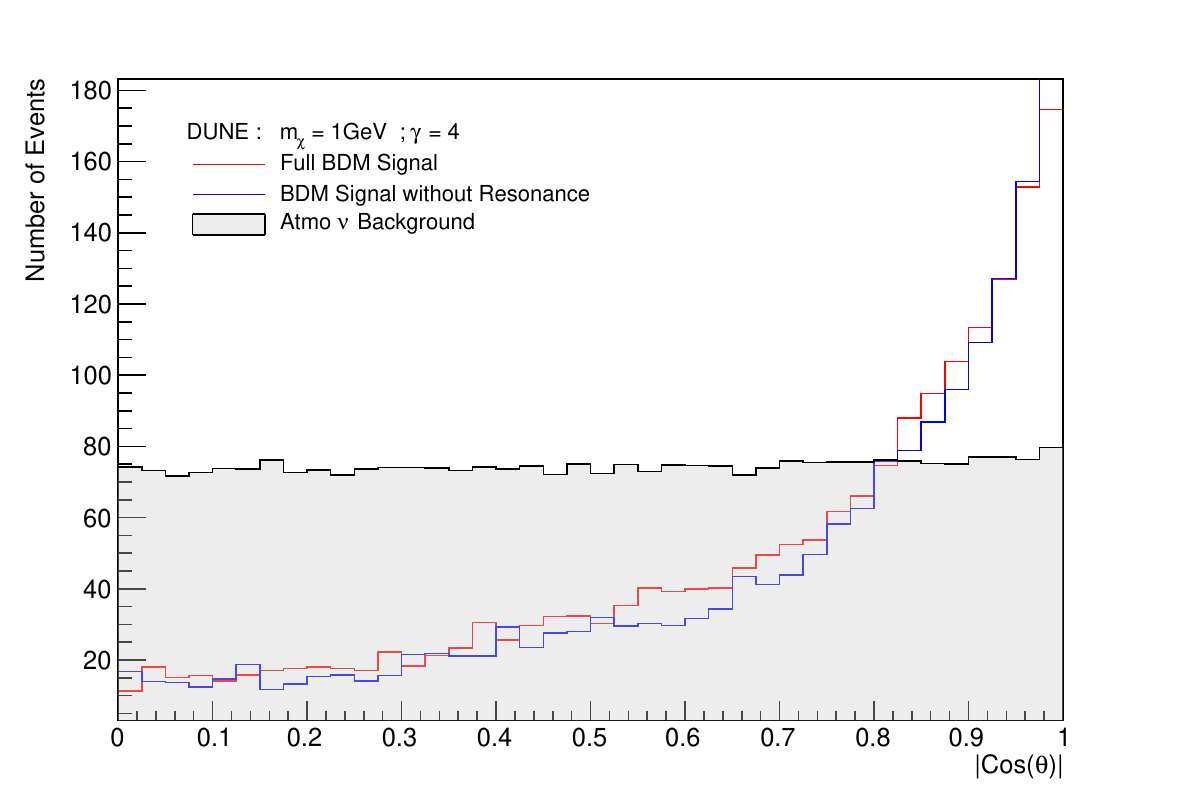}
       \end{subfigure}
       \begin{subfigure}{0.45\textwidth}
       \includegraphics[width=\linewidth]{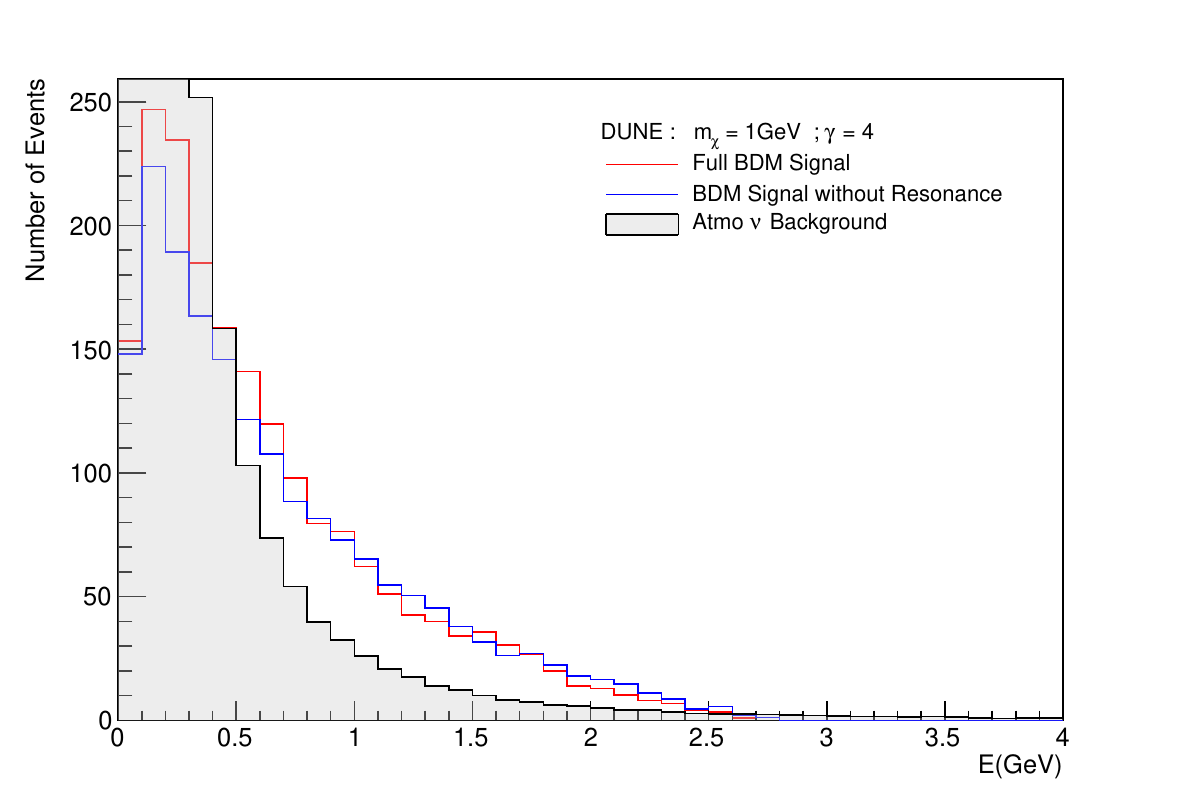}
       \end{subfigure}

       \caption{Angular (a) and energy (b) distribution for the reconstructed outgoing particles of scattering in the fermionic Dark Matter Rain model at DUNE for $m_{\chi}=1$GeV and $\gamma=4$.  The angle $\theta$ is measured with respect to the vertical.  The blue distribution shows the BDM signal excluding the contribution from resonant scattering.  The red curve shows the signal after including resonant scattering. Both BDM curves use the maximally isospin violating model. The shaded gray region shows the atmospheric neutrino background.  The normalization is arbitrary and, for the signal, depends on the particular couplings chosen.}
        \label{fig:v0_SuperK_100MeV_8Y}
    \end{figure}

We use two different selection criteria to define our search windows. Targeting lower boosts, we have a selection with $E > 0.02~{\rm GeV}$ and $\cos\theta > 0.8$.  For higher boosts, we have chosen a selection of $E > 10~{\rm GeV}$ and $\cos{\theta} > 0.9$.  The angle here is taken with respect to the expected BDM flux direction, namely the Sun in the Two Component model and the vertical in the Dark Matter Rain model.  Both selections are applied to all signal and background events.  The resulting signal and background events for each selection are compared to determine which has the greatest sensitivity.
    
Sensitivities to a signal strength $S$ are determined according to the criterion    
    \begin{equation}
        \frac{S}{\sqrt{B + \delta_B^2 B^2}} \geq 5,
    \end{equation}
where $S$ is the total number of signal events expected to pass our selection cuts at a given coupling, $B$ is the total number of atmospheric neutrino events expected to pass our selection, and $\delta_B$ is taken to be $0.3$ to account for systematic uncertainties~\cite{Acevedo:2024wmx}, though it is possible that with a sideband analysis, the systematic uncertainties could be lower on the background.  

 \begin{figure}[!tbh]
 \centering
       \includegraphics[scale=0.45]{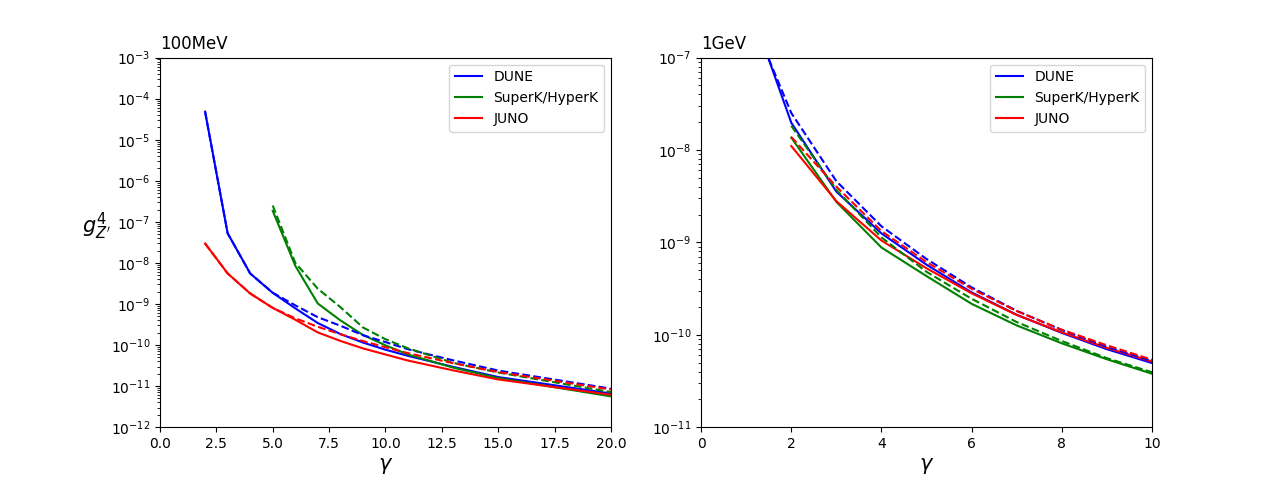}
       \caption{Expected $5\sigma$ $g_{Z^\prime}^4$ sensitivity at DUNE, Super-Kamiokande, Hyper-Kamiokande, and JUNO in the Dark Matter Rain model for masses $m_\chi = 100~{\rm MeV}$ (left) and $m_\chi= 1~{\rm GeV}$ (right). The solid lines indicate the constraints including resonant scattering in the simulation.  The dashed lines representing the solution without resonant scattering contributions.}
       \label{fig:DM_Rain_gz4}
    \end{figure}
    \begin{figure}[!tbh]
    \centering
        \includegraphics[scale=0.45]{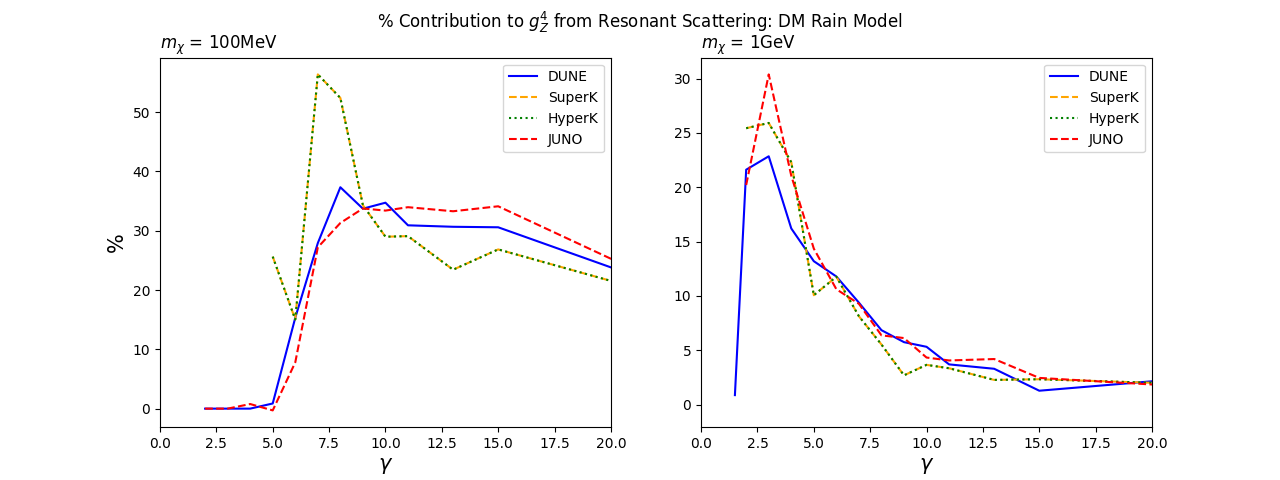}
        \caption{Percentage of resonant scattering contribution to the signal strength in the Dark Matter Rain model across various detectors for $m_\chi = 100~{\rm MeV}$ (left) and $m_\chi = 1~{\rm GeV}$ (right).}
        \label{fig:Rain_ResCon_coupling}
    \end{figure}
    \begin{figure}[!tbh]
    \centering
        \includegraphics[scale=0.45]{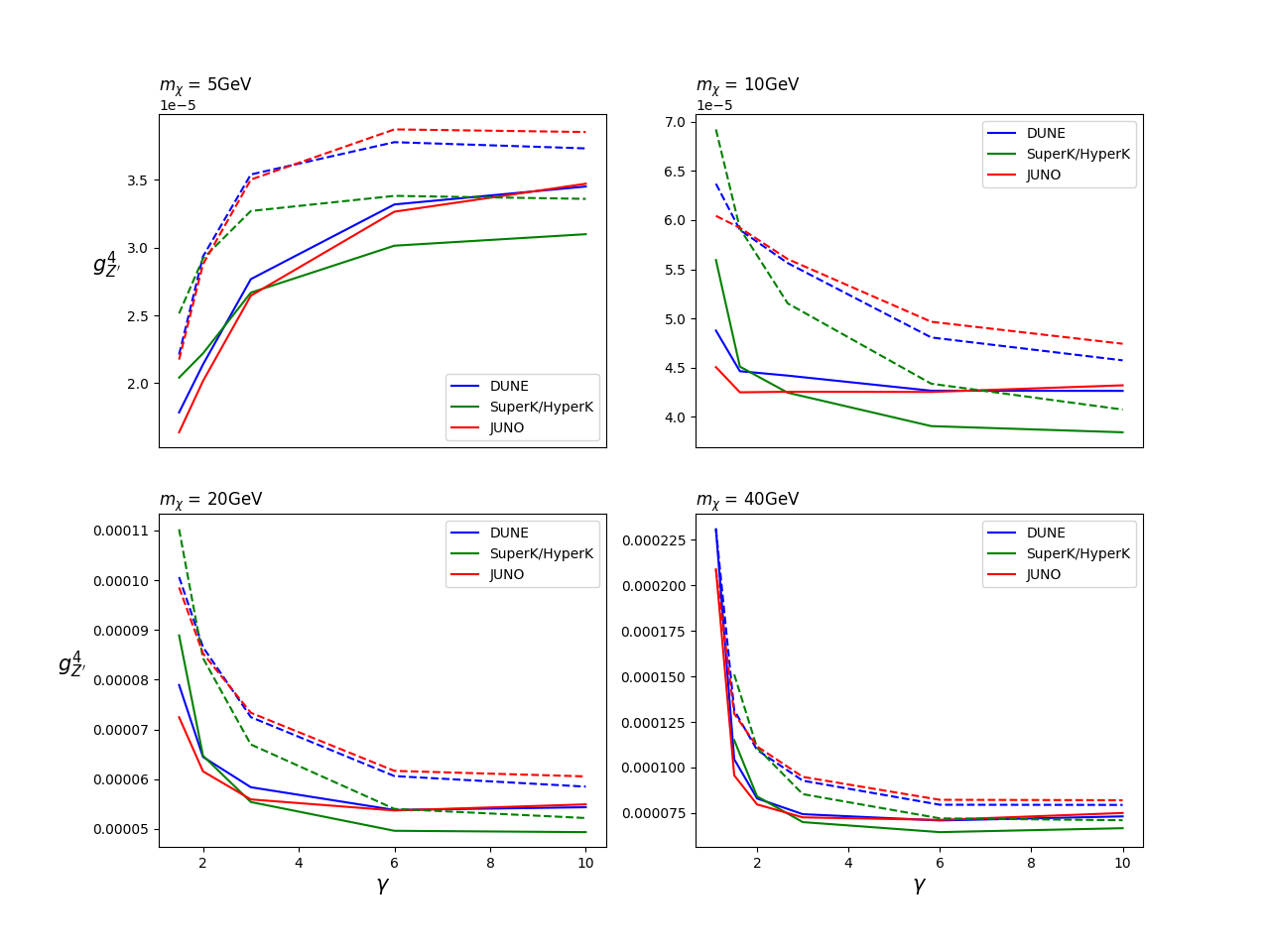}
        \caption{Same as Fig.\ \ref{fig:DM_Rain_gz4}, but for the two-component model.  The heavy dark matter scattering cross section is taken to be the maximal one compatible with bounds from direct detection.  The direct detection bounds are dominated by spin-dependent results from PICO-60L.  Masses of $m_\chi = 5~{\rm GeV}$ (top left), $m_\chi = 10~{\rm GeV}$ (top right), $m_\chi = 20~{\rm GeV}$ (bottom left), and $m_\chi = 40~{\rm GeV}$ (bottom right) are shown.}
        \label{fig:DM_Solar_gz4}
    \end{figure}
     \begin{figure}[!tbh]
     \centering
        \includegraphics[scale=0.45]{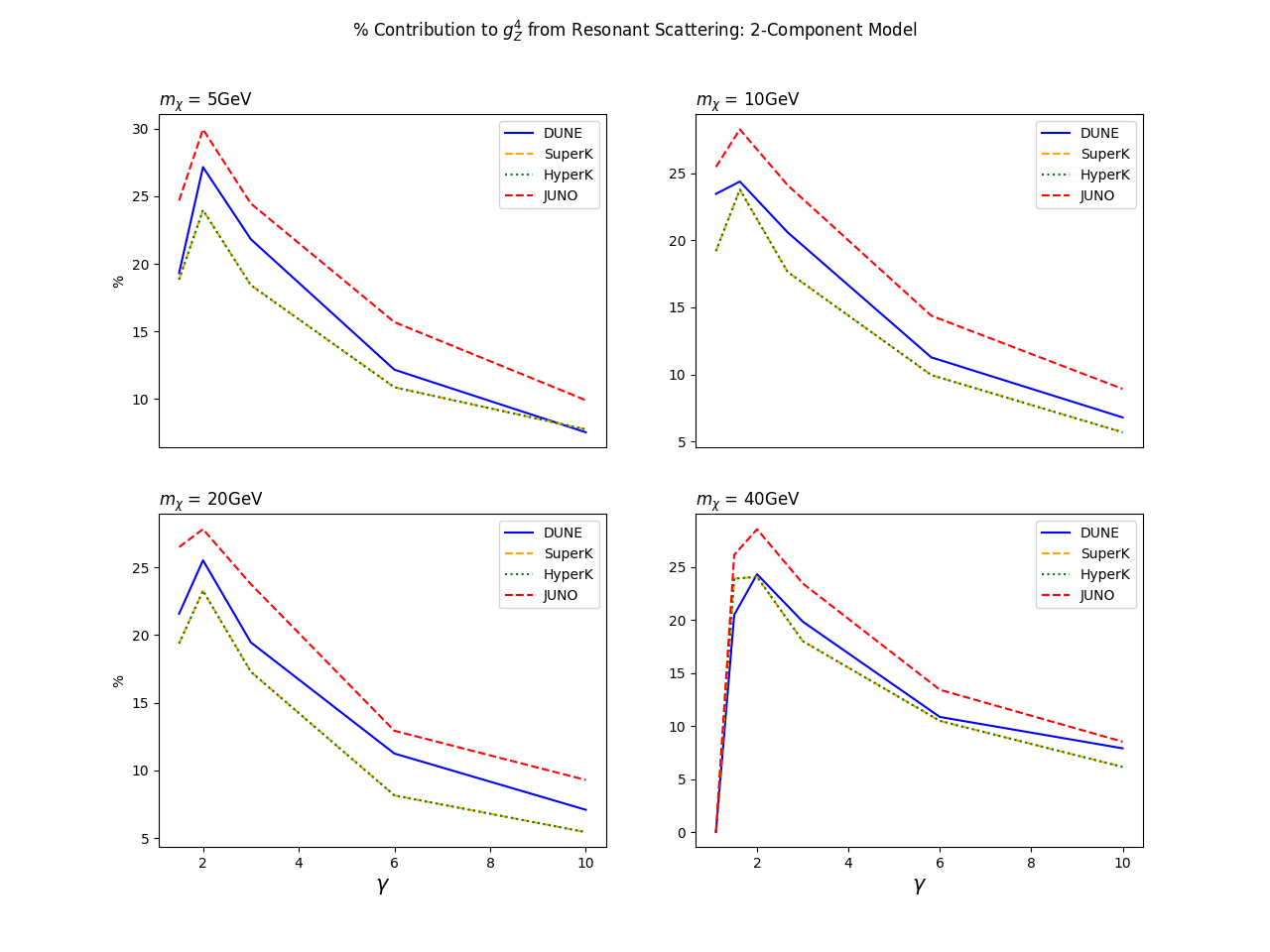}
        \caption{Same as Fig.\ \ref{fig:Rain_ResCon_coupling}, but for the Two-Component Model. Masses of $m_\chi = 5~{\rm GeV}$ (top left), $m_\chi = 10~{\rm GeV}$ (top right), $m_\chi = 20~{\rm GeV}$ (bottom left), and $m_\chi = 40~{\rm GeV}$ (bottom right) are shown.}
        \label{fig:Solar_ResCon}
    \end{figure}
The resulting sensitivities are shown in Figs.\ \ref{fig:DM_Rain_gz4} and \ref{fig:DM_Solar_gz4} for the Dark Matter Rain and Two-Component models respectively.  In addition, we determine the extra sensitivity due to the inclusion of resonant scattering.  This increase in sensitivity to $g_{Z'}^4$ is shown in Figs.\ \ref{fig:Rain_ResCon_coupling} and \ref{fig:Solar_ResCon} for the Dark Matter Rain and Two-Component models respectively.  

\begin{figure}[!tbh]
\centering
        \includegraphics[width=10cm]{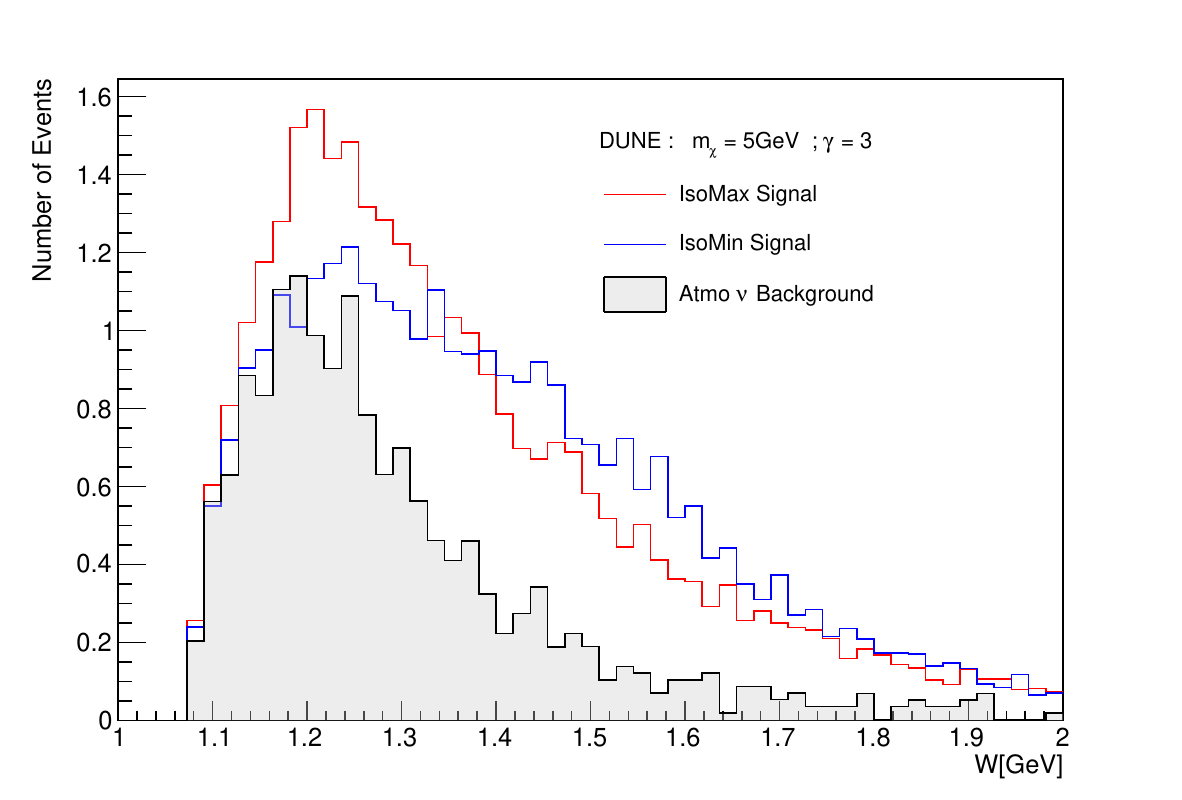}
        \caption{The invariant mass of the reconstructed final state particles for events which include only single proton and charged pion production, above the energy thresholds for DUNE with 10yr exposure time and normalized to a number of events corresponding to a 5$\sigma$ discovery. Both BDM curves are generated for scalar DM in the two-component model at a mass of $5~\text{GeV}$ and a boost factor of 3. The red (blue) line indicates the signal from events generated with maximal (no) isospin symmetry violation.}
        \label{fig:W_5eV_3Y}
    \end{figure}

\newpage
\section{Discussion and Outlook}
\label{sec:discussion}

This work completes the simulation of boosted dark matter scattering processes of nuclei mediated by a vector.  We plan to request that this additional scattering module be incorporated into the \lstinline{GENIE} code base so that it may be integrated into experimental simulation pipelines.  It will be important to further determine the extent to which baryonic resonances can be resolved with a more complete simulation of proton and pion reconstruction.  For example, within the context of a detector such as DUNE, it could be somewhat challenging to reconstruct the vertically oriented tracks expected in the dark matter rain model.

With the development of the new ACHILLES generator~\cite{Isaacson:2022cwh}, in which it should be possible to incorporate BSM models via FeynRules-generated~\cite{Alloul:2013bka} model files, it will be possible to compare the results of the two generators for validation.  Furthermore, this comparison will allow for the assessment of uncertainties in the boosted dark matter prediction.

In addition to the validation of the boosted dark matter models considered in this work, there are several variants that are of interest and could be incorporated into \lstinline{GENIE} in future work.  For example, the dark matter could scatter by a scalar or pseudo-scalar mediator or an excitable dark sector state could be excited~\cite{Kim:2016zjx}.  We leave the detailed study of these alternatives to future work.

\section*{Acknowledgments}

JB would like to thank Yun-Tse Tsai for helpful discussions.  This material is based upon work supported by the National Science Foundation under Grant No.\ 2413017.  This work was performed in part at the Aspen Center for Physics, which is supported by National Science Foundation grant PHY-2210452.

\appendix
\section{Resonant Scattering Kinematics}\label{sec:resonant-kinematics}
The production cross section of a resonance $\mathcal{N}^*$ of mass $M$ in a $2 \to 2$ process can be written in the form
\begin{equation}\label{eq:cross-section-general}
\left(\frac{d^2 \sigma}{d q^2 d W}\right)_{\text{CM}} = \frac{1}{16 \pi (1 - m_\chi^2/E^2)}\frac{W}{m_\mathcal{N}}\frac{1}{2m_\mathcal{N}  E^2} \overline{|\mathcal{M}|^2} \delta(W^2 - M^2)
\end{equation}
where $W$ is the invariant mass of the hadronic part of the final state, $E$ and $m_\chi$ are the energy and mass of the boosted dark matter, and $m_\mathcal{N}$ is the mass of the initial nucleon.  We include an explicit delta function to allow for eventual smearing by a Breit-Wigner propagator. In the center of mass frame (CM) we have $p_1 + p_2 = p_3 + p_4$.

The squared amplitude is written in terms of a ``leptonic'' tensor $L$ and a hadronic tensor $W$ as
\begin{equation}
|T(\chi \mathcal{N} \rightarrow \chi \mathcal{N}^*)|^2 = \bar{|\mathcal{M}|^2} \delta(W^2 - M^2)] = \frac{g_{Z^\prime}^4}{(q^2 - m_{Z^\prime}^2)^2} L_{\alpha \beta} W^{\alpha \beta}.
\end{equation}
Substituting this expression into \eqref{eq:cross-section-general}, we can write the differential cross section in the convenient form
\begin{equation}
\left(\frac{d^2 \sigma}{d q^2 d W}\right)_{\text{CM}} = \sigma_0 \left( \frac{Q^2}{-q^2 E^2}\frac{1}{ 2m_\mathcal{N}} L_{\alpha \beta}  W^{\alpha \beta} \right),
\end{equation}
where
\begin{equation}
\sigma_0 = \frac{g_{Z^\prime}^4}{16 \pi (1 - m_\chi^2/E^2) (q^2 - m_{Z^\prime}^2)^2} \frac{W}{m_\mathcal{N}} \frac{-q^2}{Q^2}.
\end{equation}

The leptonic current can be calculated by spin averaging the dark matter currents.  For fermionic DM, the result is 
\begin{multline}
L_{\alpha \beta} = g_{\alpha \beta} \left( ( Q_x-Q_+)m_{\chi}^2 +  \frac{q^2}{2}Q_+\right)
+  q_\alpha q_\beta (Q_x - Q_+)m_\chi^2\left( \frac{ q^2-2m_{Z^\prime}^2}{m_{Z^\prime}^4} \right)\\
+Q_+\left(p_{1\alpha}p_{3\beta} + p_{1\beta}p_{3\alpha}\right)+ Q_-\left( i \epsilon_{\alpha \beta p_1 p_3} \right)
\end{multline}

where, in terms of the left- ($Q_L^\chi$) and right-handed ($Q_R^\chi$) dark matter charges, $Q_+ =  (Q_L^\chi)^2 + (Q_R^\chi)^2$, $Q_- =  (Q_L^\chi)^2 - (Q_R^\chi)^2$ and $Q_x = 2 Q_L^\chi Q_R^\chi$. Here, $p_1$ and $p_3$ are the incoming and outgoing lepton 4-momenta. For scalar DM, the tensor is written simply as
\begin{equation}
L_{\alpha \beta} = (Q_S^\chi)^2(p_1 + p_3)_\alpha (p_1 + p_3)_\beta,
\end{equation}
where $Q_S^\chi$ is the scalar dark matter charge. 

The hadronic current can be written in the conventional form
\begin{equation}
W^{\alpha\beta} = -g^{\alpha\beta} W_1 + p_2^\alpha p_2^\beta W_2 +  q^\alpha q^\beta W_3 + (p_2^\alpha q^\beta + q^\alpha p_2^\beta) W_4 - i \epsilon^{\alpha\beta p_2 q} W_5
\end{equation}
Where, $p_2$ and $p_4$ are the incoming and outgoing hadron 4-momenta respectively.
The intermediate vector boson polarization can be decomposed into its off-shell polarizations $e^\mu$ in the resonance rest frame (RRF). In the work of Rein and Sehgal~\cite{Rein:1980wg}, the lepton mass is taken to be negligible such that all leptonic currents are conserved and $e\cdot q = 0$ resulting in three off-shell polarizations. Since the dark matter mass cannot be neglected in general, a fourth independent polarization $e_q^\mu$ needs to be included. The 3-momentum transfer of the virtual boson is taken along the z-axis in the RRF so that $q^\mu = (\nu^*;0,0,Q^*)$. Throughout this appendix, starred variables refer to RRF quantities.  We take the four polarizations to be
\begin{subequations}\label{eq:polarization-vectors-RRF}
    \begin{equation}
        e_S^\mu = \sqrt{\frac{1}{-q^2}}(Q^*;0,0,\nu^*)
    \end{equation}
    \begin{equation}
        e_R^\mu = \sqrt{\frac{1}{2}}(0;-1,-i,0)
    \end{equation}
    \begin{equation}
        e_L^\mu = \sqrt{\frac{1}{2}}(0;+1,-i,0)
    \end{equation}
    \begin{equation}
        e_q^\mu = \sqrt{\frac{1}{-q^2}}(-\nu^*;0,0,-Q^*)
    \end{equation}
\end{subequations}
The three-momentum transfer in the RRF and the lab frame are related by ~\cite{Ravndal:1973xx,Rein:1980wg}
\begin{equation}
    Q^* = Q \frac{m_\mathcal{N}}{M}
\end{equation}

The hadronic tensor can then be expanded in terms of the polarization vector basis,
\begin{equation}
\sigma_i = e_{i}^{\alpha *} e_{i}^{\beta} W_{\alpha \beta} 
\end{equation}

The polarization structure factors are related to the hadronic tensor form factors by
\begin{subequations}
    \begin{equation}
        \sigma_L = W_1 - M Q^* W_5;
    \end{equation}
     \begin{equation}
        \sigma_R = W1 + M Q^* W_5;
    \end{equation}
     \begin{equation}
        \sigma_S = -W1 - \frac{M^2 Q^{*2}}{q^2}W_2;
    \end{equation}
    \begin{equation}
        \sigma_q = W_1 - \frac{M^2 \nu^{*2}}{q^2}W_2 - q^2 W_3 - 2\nu^* M W_4.
    \end{equation}
\end{subequations}
From this, we find that the hadornic tensor is given by

\begin{equation}
\begin{gathered}
W^{\alpha\beta} = \sigma_L \left( -\frac{g^{\alpha \beta}}{2} - \frac{q^2}{2 m_{\mathcal{N}}^2 Q^2}p_2^\alpha p_2^\beta - \frac{- i \epsilon^{\alpha\beta p_2 q}}{2 m_{\mathcal{N}} Q} + \frac{\nu^2 + Q^2}{q^2 Q^2}q^\alpha q^\beta  \right)\\
+\sigma_R \left( -\frac{g^{\alpha \beta}}{2} - \frac{q^2}{2 m_{\mathcal{N}}^2 Q^2}p_2^\alpha p_2^\beta + \frac{- i \epsilon^{\alpha\beta p_2 q}}{2 m_{\mathcal{N}} Q} + \frac{\nu^2 + Q^2}{q^2 Q^2}q^\alpha q^\beta  \right)\\
+\sigma_S \left(\frac{\nu^2}{q^2 Q^2}q^\alpha q^\beta - \frac{q^2}{m_{\mathcal{N}}^2 Q^2}p_2^\alpha p_2^\beta \right)
+\sigma_Z \left(\frac{1}{-q^2}q^\alpha q^\beta \right)
\end{gathered}
\end{equation}

Working in terms of the kinematic variables,
\begin{subequations}
    \begin{equation}
        u = \frac{E + E' + Q}{2E}
    \end{equation}
    \begin{equation}
        v = \frac{E+E'-Q}{2E}
    \end{equation}
\end{subequations}
the leptonic and hadronic tensors can be directly contracted to give the partial cross-section in terms of cross-section structure factors and kinematic terms. For fermion DM the result can be expressed as
\begin{multline}
L_{\alpha \beta}W^{\alpha \beta} = \left( \frac{-q^2 E^2}{Q^2}\right) [\\
\sigma_L \left((Q_L^\chi)^2 u^2 
+ (Q_R^\chi)^2 v^2
+\frac{2m_\chi^2 Q^2 Q_L^\chi Q_R^\chi}{ E^2 q^2}\right)\\
\sigma_R \left( (Q_L^\chi)^2 v^2 
+(Q_R^\chi)^2 u^2
+\frac{2m_\chi^2 Q^2 Q_L^\chi Q_R^\chi}{ E^2 q^2}\right)\\
\sigma_S \left( 2 u v ((Q_L^\chi)^2 + (Q_R^\chi)^2)
+\frac{m_\chi^2 Q^2 (Q_L^\chi - Q_R^\chi)^2}{ E^2 q^2}\right)\\
\sigma_q \left( - \frac{m_\chi^2 Q^2 (Q_L^\chi - Q_R^\chi)^2 (m_{Z'}^2 - q^2)^2 }{E^2 m_{Z'}^4 q^2} \right) ].
\end{multline}
For scalar DM, the result is
\begin{equation}
L_{\alpha \beta}W^{\alpha \beta} = \left( \frac{-q^2 E^2}{Q^2}\right)
(Q_S^\chi)^2 [ (\sigma_L + \sigma_R) \left(2 u v + \frac{2m_\chi^2 Q^2}{q^2 E^2} \right) + \sigma_S (u + v)^2 ]
\end{equation}

The differential cross section can then be expressed in the usual form as
\begin{equation}
    \left(\frac{d^2 \sigma}{d q^2 d W}\right)_{\text{CM}} =
    \sigma_0\frac{1}{ 2m_\mathcal{N}}\left[ \sigma_L A_L + \sigma_R A_R + \sigma_S A_S + \sigma_q A_q \right]
\end{equation}
where, the variables $A_i$ depend on the nature of the DM. For fermion DM,
\begin{subequations}
    \begin{equation}
        A_L = (Q_L^\chi)^2 u^2 
+ (Q_R^\chi)^2 v^2
+\frac{2m_\chi^2 Q^2 Q_L^\chi Q_R^\chi}{ E^2 q^2};
    \end{equation}
    \begin{equation}
        A_R = (Q_L^\chi)^2 v^2 
+(Q_R^\chi)^2 u^2
+\frac{2m_\chi^2 Q^2 Q_L^\chi Q_R^\chi}{ E^2 q^2};
    \end{equation}
    \begin{equation}
        A_S =  2 u v ((Q_L^\chi)^2 + (Q_R^\chi)^2)
+\frac{m_\chi^2 Q^2 (Q_L^\chi - Q_R^\chi)^2}{ E^2 q^2};
    \end{equation}
    \begin{equation}
        A_q = - \frac{m_\chi^2 Q^2 (Q_L^\chi - Q_R^\chi)^2 (m_{Z'}^2 - q^2)^2 }{E^2 m_{Z'}^4 q^2}.
    \end{equation}
\end{subequations}
For scalar DM,
\begin{subequations}
    \begin{equation}
        A_L = A_R = (Q_S^\chi)^2\left( 2 u v + \frac{2m_\chi^2 Q^2}{q^2 E^2} \right);
    \end{equation}
    \begin{equation}
        A_S = (Q_S^\chi)^2\left( (u + v)^2\right);
    \end{equation}
    \begin{equation}
        A_q = 0.
    \end{equation}
\end{subequations}

The helicity structure factors are determined from the hadronic current, expressed in terms of the resonance production matrix elements in the RRF. The hadronic current operator is conventionally written with the resonance mass factored out as $2MF_\mu$ \cite{Rein:1980wg}. Projected along the polarization basis the helicity structure factors, ignoring the spin-average, are equivlent to,
\begin{equation}
\begin{gathered}
e^{\alpha *}e^{\beta} W_{\alpha \beta} = (2M)^2 \langle \mathcal{N}|e^{\alpha*}\cdot F_\alpha|\mathcal{N}^*\rangle \langle \mathcal{N}^*|e^{\beta}\cdot F_\beta|\mathcal{N}\rangle \frac{\delta(W - M)}{2M}\\ = (2m_\chi)\frac{M}{m_\chi} |\langle \mathcal{N}|e^{\alpha*}\cdot F_\alpha|\mathcal{N}^*\rangle|^2\delta(W - M),
\end{gathered}
\end{equation}
The general hermitian operator, $F_\mu$, may be decomposed in terms of a vector and axial-vector component as,
\begin{equation}
    F_\mu = F^V_\mu - F^A_\mu = \frac{1}{2M}(V_\mu - A_\mu),
\end{equation}
here the minus sign is conventional ~\cite{Ravndal:1973xx,Rein:1980wg}.

The hadronic martix elements $V_\mu$ and $A_\mu$ are determined in the RRF by the FKR model~\cite{Feynman:1971wr}. By assumption, the target nucleon $\mathcal{N}$ is in the ground state with spin $j_z^{\prime} = \pm \frac{1}{2}$. Helicity amplitudes can then be represented as,

\begin{subequations}\label{eq:helicity-amplitudes}
    \begin{equation}
        f_{\pm|2j_z|} = \langle \mathcal{N},j_z \pm 1 |F_\pm|\mathcal{N}^*,j_z \rangle
    \end{equation}
    \begin{equation}
        f_{[0,D]\pm} = \langle \mathcal{N}, \pm \frac{1}{2}|F_{[0,D]}|\mathcal{N}^*,\pm \frac{1}{2} \rangle,
    \end{equation}
\end{subequations}

where $F_\pm$, $F_{0\pm}$ and $F_{D\pm}$ are found by  projecting $F_\mu = (F_t,F_x,F_y,F_z)$, along the polarization basis vectors in terms of RRF quantities \cite{Rein:1980wg}. Here, we used the partialy conserved axial current (PCAC) relation
\begin{equation}
    q^\mu F_\mu = q^\mu F^V_\mu - q^\mu F^A_\mu = 0 - D
\end{equation}
to define the total divergence $D=-q^\mu F_\mu=\frac{1}{2M}q^\mu A_\mu$ and to write $F_z$ in terms of $F_t$ and $D$ with $F_t = e_0^\mu F_\mu$ by letting $e_0^\mu = (1,0,0,0)$.

Through a redefinition of the structure factors, we find the differential cross section is
\begin{equation}
    (\frac{d^2 \sigma}{d q^2 d W})_{\text{CM}} =\sigma_0\left[ \sigma_L A_L + \sigma_R A_R + \sigma_S A_S + \sigma_q A_q \right],
\end{equation}
where the helicity structure factors are now defined, including the spin average, as
\begin{subequations}
    \begin{equation}
         \sigma_{L}=\frac{M}{m_{\mathcal{N}}}\frac{1}{2}(|f_{+3}|^2 + |f_{+1}|^2) \delta (W - M),
    \end{equation}
    \begin{equation}
        \sigma_{R}=\frac{M}{m_{\mathcal{N}}}\frac{1}{2}(|f_{-3}|^2 + |f_{-1}|^2) \delta (W - M),
    \end{equation}
    \begin{equation}
        \sigma_{S}=\frac{m_{\mathcal{N}}}{M}\frac{Q^{2}}{-q^2}\frac{1}{2}(|f_{0+}|^2 + |f_{0-}|^2) \delta (W - M),
    \end{equation}
    \begin{equation}
        \sigma_{q}=\frac{m_{\mathcal{N}}}{M}\frac{Q^{2}}{-q^2}\frac{1}{2}(|f_{D+}|^2 + |f_{D-}|^2) \delta (W - M),
    \end{equation}
\end{subequations}
with helicity amplitudes given by \eqref{eq:helicity-amplitudes}. The operators for the helicity amplitudes, expressed in the RRF, are written in terms of the FKR helicity operators as,
    \begin{subequations}
        \begin{equation}
            F_+ = \frac{e_R^\mu V_\mu }{2M} - \frac{e_R^\mu A_\mu}{2M}
        \end{equation}
        \begin{equation}
            F_- = \frac{e_L^\mu V_\mu }{2M} - \frac{e_L^\mu A_\mu}{2M}
        \end{equation}
        \begin{equation}
            F_0  = \frac{-q^2}{Q^{*2}}\frac{e_0^\mu V_\mu}{2M} - \frac{-q^2}{Q^{*2}}\frac{e_0^\mu A_\mu}{2M} + \frac{\nu^*}{Q^{*2}} \frac{q^\mu A_\mu}{2M}
        \end{equation}
        \begin{equation}
            F_D = \frac{Q^*}{Q^{*2}} \frac{q^\mu A_\mu}{2M}.
        \end{equation}
    \end{subequations}
Projecting the FKR operators along the polarization basis, we can write the operators for the helicity amplitudes in a simplified form. The reduction of $F_\pm$ and $F_0$ has been carried out in previous work \cite{Ravndal:1973xx,Rein:1980wg} with results,
\begin{subequations}
    \begin{equation}
        F_\pm^V = -9e_a(R^V\sigma_\pm + T^V a_\mp)e^{-\lambda a_z},
    \end{equation}
    \begin{equation}
        F_\pm^A = \pm9e_a^\prime(R^A\sigma_\pm + T^A a_\mp)e^{-\lambda a_z},
    \end{equation}
    \begin{equation}
        F_0^V = 9e_a S e^{-\lambda a_z},
    \end{equation}
    \begin{equation}
        F_0^A = -9e_a^\prime [C\sigma_z + B(\vec{\sigma}\cdot\vec{a})]e^{-\lambda a_z},
    \end{equation}
\end{subequations}
where,
\begin{eqnarray}
\lambda &=& \sqrt{\frac{2}{\Omega}}\frac{m_\mathcal{N}}{M}Q\\
T^V &=& \frac{G^V}{3M}\sqrt{\frac{\Omega}{2}}\\
T^A &=& \frac{2}{3}\frac{G^A Z m_\mathcal{N}}{M}\sqrt{\frac{\Omega}{2}} \frac{ Q }{(M+m_{\mathcal{N}})^2 -q^2}\\
R^V &=&  \sqrt{2} G^V \frac{m_\mathcal{N}}{M}Q(\frac{(M+m_\mathcal{N})}{(M+m_\mathcal{N})^2 -q^2})\\
R^A &=& \frac{\sqrt{2} G^A Z}{6M}[(M+m_\mathcal{N})+\frac{2 N\Omega M}{(M+m_{\mathcal{N}})^2 -q^2}]\\
S &=& (\frac{-q^2}{Q^{2}})\frac{G^V}{6m_\mathcal{N}^2}[3Mm_\mathcal{N} + q^2 - m_\mathcal{N}^2 ]\\
C &=& \frac{Z G^A}{6m_\mathcal{N}Q}
( M^2 - m_\mathcal{N}^2 + N\Omega\frac{M^2 - m_\mathcal{N}^2 + q^2}{(M+m_{\mathcal{N}})^2 -q^2}) \\
B &=& \frac{Z G^A}{3M}\sqrt{\frac{\Omega}{2}}(1+ \frac{M^2 - m_\mathcal{N}^2 + q^2}{(M+m_{\mathcal{N}})^2 -q^2})\\
\end{eqnarray}
$\Omega = 1.05 (GeV)^2$ is the harmonic oscillator constant and $N$ is the number of excitations in $\mathcal{N}^*$. $G^V$,$G^A$ and $Z$ remain undetermined. 

The undetermined operator, $F_D$, has only axial components and may be expressed in the same form as $F_0^A$ with new coefficients as,
\begin{equation}
    F_D = -9e_a^\prime [C_D\sigma_z + B_D(\vec{\sigma}\cdot\vec{a})]e^{-\lambda a_z}.
\end{equation}
The coefficients here may be expressed in terms of $C$ and $B$ or solved for directly using the divergence expressed in \cite{Feynman:1971wr,Ravndal:1973xx}.The result is the same,
\begin{subequations}
    \begin{equation}
        C_D = \frac{ZG_A}{3}\left(1 + \frac{3q^2 + N\Omega}{(M+m_\mathcal{N})^2 - q^2}\right)
    \end{equation}
    \begin{equation}
        B_D = \frac{ZG_A}{2m_\mathcal{N}Q}\sqrt{\frac{\Omega}{2}}\left(M - m_\mathcal{N} \right)
    \end{equation}
\end{subequations}

There are three undetermined constants: $G^V,G^A,Z$. The transition form factors are assumed to have the form:\\
\begin{equation}
G^{V,A}(q^2) = (1 - \frac{q^2}{4 m_{\mathcal{N}}^2})^{\frac{1}{2} - N}(\frac{1}{1-q^2 / m_{V,A}^2})^2
\end{equation}
The constants $Z,m_V,m_A$ are experimentally determined. Their default values in \lstinline{GENIE} are $Z=0.76338$, $m_V=0.840$, $m_A=1.120$~\cite{andreopoulos2015genieneutrinomontecarlo}. \\

All helicity production amplitudes for resonance multiplets can be determined by the FKR model \cite{Feynman:1971wr} for the operators $\{e_a;\sigma_a;a\}$. Which represent the unitary-spin dependence, spin dependence and radial dependence for the bound state of a quark triplet.\\
The radial and spin operator amplitudes are tabulated or calculated in the previous work by FKR \cite{Feynman:1971wr} and will not need to be adjusted for the DM case. The charge operator, $e_a$, is taken to be the charge on the first quark in the triplet. Here, $e_a$ and $e'_a$ must be different operators since the precise interaction of the DM is unknown.
\begin{equation}
e_a = \frac{Q_u^V - Q_d^V}{2}\tau_3 + \frac{Q_u^V + Q_d^V}{2}I
\end{equation}
\begin{equation}
e'_a = \frac{Q_u^A - Q_d^A}{2}\tau_3 + \frac{Q_u^A + Q_d^A}{2}I
\end{equation}
In this form, the substitutions $Q_u^V \rightarrow \frac{2}{3}$ and $Q_d^V \rightarrow -\frac{1}{3}$ with $Q_u^A=Q_d^A=0$ reproduce the results for resonance production mediated by EM interactions. The couplings $Q_u^V \rightarrow \frac{1}{2} - 2x\frac{2}{3}$ ; $Q_d^V \rightarrow -\frac{1}{2} - 2x\frac{-1}{3}$ ; $Q_u^A \rightarrow \frac{1}{2}$ ; $Q_d^A \rightarrow -\frac{1}{2}$, where $x = \sin^2\theta_W$ and $\theta_W$ is the weak mixing angle, produce the results for resonance production via neutral current interactions.  This agreement has been verified by comparing with the results of Ref.\ \cite{Rein:1980wg}.

To consider finite width effects, the delta function can be replaced in the narrow-width approximation by a Breit-Wigner factor,
\begin{equation}
    \delta(W- M) \rightarrow \frac{1}{2\pi} \frac{\Gamma}{(W-M)^2 + \Gamma^2/4}
\end{equation}

\newpage

\section{Helicity Amplitude Tables}\label{sec:helicity-amp}
To simplify expressions in the table, the following notation is used.
\begin{subequations}
    \begin{equation}
    R^{\pm} = R^{\pm}_u -R^{\pm}_d
    \end{equation}
    \begin{equation}
    T^{\pm} = T^{\pm}_u - T^{\pm}_d
    \end{equation}
    \begin{equation}
        Q^A = Q^A_u - Q^A_d
    \end{equation}
    \begin{equation}
        Q^V = Q^V_u - Q^V_d
    \end{equation}
    where,
    \begin{equation}
    R_q^\pm = R^VQ_q^V \pm R^AQ_q^A
    \end{equation}
    \begin{equation}
    T_q^\pm = R^VQ_q^V \pm R^AQ_q^A
    \end{equation}
\end{subequations}
Additionally, as a shorthand notation for other combinations of quark charges and coefficients,
\begin{subequations}
    \begin{equation}
    Q_{C_1 q\pm C_2 q'}^{(V,A)} = C_1 Q_q^{(V,A)} \pm C_2 Q_{q'}^{(V,A)}
    \end{equation}
    \begin{equation}
    R^{(\pm)}_{C_1 q\pm C_2 q'} = C_1 R^{(\pm)}_q \pm C_2 R^{(\pm)}_{q'}
    \end{equation}
    \begin{equation}
    T^{(\pm)}_{C_1 q\pm C_2 q'} = C_1 T^{(\pm)}_q \pm C_2 T^{(\pm)}_{q'}
    \end{equation}
\end{subequations}

where, $C_1$ and $C_2$ are numerical coefficients. Note that, $Q_{u-d}^V = Q^V$ and $Q_{u-d}^A = Q^A$, similarly for $R^\pm$ and $T^\pm$. Both $Q^V$ and $Q^A$ will vanish if the quark charges are aligned in the vector axial-vector basis. 

Lastly, we note that the coefficient $B$, and similarly $B_D$, always appear in combination with $C\lambda$, or $C_D\lambda$, as $NB-C\lambda$, or $NB_D - C_D\lambda$, where $N$ is just a number. We also define
\begin{subequations}
    \begin{equation}
        \mathcal{B}(N) = NB-C\lambda
    \end{equation}
    \begin{equation}
        \mathcal{B}_D(N) = NB_D-C_D\lambda
    \end{equation}
\end{subequations}

\newpage
{\small
\begin{longtable}[c]{|p{2.4cm}|p{0.1cm}|p{4.8cm}|p{4.8cm}|}

\hline
\multicolumn{2}{|c|}{Resonance Production}&
\multicolumn{2}{|c|}{Nucleon $\mathcal{N}$}\\
\hline
$\mathcal{N}^*$ &  & p & n\\
\hline
\endfirsthead

 \hline 
 \multirow{8}{1em}{$P_{33}(1234)$\\$^4(10)_{3/2}[56,0^+]_0$}&
 $f_{-3}$ & $\sqrt{6} R^-$ &	$\sqrt{6} R^-$\\
&$f_{-1}$ &$\sqrt{2} R^-$&	 $\sqrt{2} R^-$\\
&$f_{+1}$ &$\sqrt{2} R^+$ &	$\sqrt{2} R^+$\\
&$f_{+3}$ &$\sqrt{6} R^+$ &	$\sqrt{6} R^+$\\
&$f_{0+}$ &$2 C \sqrt{2} Q^A$&	$2 C \sqrt{2} Q^A$\\
&$f_{0-}$ &$2 C \sqrt{2} Q^A$&	$2 C \sqrt{2} Q^A$\\
&$f_{D+}$ &$2 C_D \sqrt{2} Q^A$&	$2 C_D \sqrt{2} Q^A$\\
&$f_{D-}$ &$2 C_D \sqrt{2} Q^A$&	$2 C_D \sqrt{2} Q^A$\\
\hline
\hline
\multirow{8}{1em}{$S_{11}(1540)$\\$^2(8)_{1/2}[70,1^-]_1$}&
$f_{-3}$ & 0 & 0\\
&$f_{-1}$ &$-\frac{1}{2\sqrt{3}}(6T^- +\lambda\sqrt{2}R^-_{5u+d})$ &
$\frac{1}{2\sqrt{3}}(6T^- -\lambda\sqrt{2}R^-_{u+5d})$\\
&$f_{+1}$ &$-\frac{1}{2\sqrt{3}}(6T^+ +\lambda\sqrt{2}R^+_{5u+d})$ & 
$\frac{1}{2\sqrt{3}}(6T^+-\lambda\sqrt{2}R^+_{u+5d})$\\
&$f_{+3}$ &0 & 0\\
&$f_{0+}$ &$-\frac{1}{\sqrt{6}}(Q_{5u+d}^A\mathcal{B}(3)- 3 \lambda S Q^V)$ & 
 $-\frac{1}{\sqrt{6}}(Q_{u+5d}^A\mathcal{B}(3)+3 \lambda S Q^V)$\\
&$f_{0-}$ & $-\frac{1}{\sqrt{6}}(Q_{5u+d}^A\mathcal{B}(3) + 3 \lambda S Q^V)$ & 
 $-\frac{1}{\sqrt{6}}(Q_{u+5d}^A\mathcal{B}(3)-3 \lambda S Q^V)$\\
&$f_{D+}$ &$-\frac{1}{\sqrt{6}}Q_{5u+d}^A \mathcal{B}_D(3)$ &	
$-\frac{1}{\sqrt{6}}Q_{u+5d}^A \mathcal{B}_D(3)$\\
&$f_{D-}$ &$-\frac{1}{\sqrt{6}}Q_{5u+d}^A \mathcal{B}_D(3)$ &	
$-\frac{1}{\sqrt{6}}Q_{u+5d}^A \mathcal{B}_D(3)$\\
\hline
\multirow{8}{1em}{$D_{13}(1525)$\\$^2(8)_{3/2}[70,1^-]_1$}&
$f_{-3}$ &$-\frac{3}{\sqrt{2}}T^-$&
$\frac{3}{\sqrt{2}}T^-$\\
&$f_{-1}$ &$-\sqrt{\frac{3}{2}}T^- +\frac{\lambda}{\sqrt{3}}R^-_{5u+d}$&
$\sqrt{\frac{3}{2}}T^- +\frac{\lambda}{\sqrt{3}}R^-_{u+5d}$\\
&$f_{+1}$ &$\sqrt{\frac{3}{2}}T^+ -\frac{\lambda}{\sqrt{3}}R^+_{5u+d}$&
$-\sqrt{\frac{3}{2}}T^+ -\frac{\lambda}{\sqrt{3}}R^+_{u+5d}$\\
&$f_{+3}$ &$\frac{3}{\sqrt{2}}T^+$&
$-\frac{3}{\sqrt{2}}T^+$\\
&$f_{0+}$ &$-\frac{\lambda}{\sqrt{3}}(C Q_{5u+d}^A+3 S Q^V)$&
$-\frac{\lambda}{\sqrt{3}}(C Q_{u+5d}^A-3 S Q^V)$\\
&$f_{0-}$ &$\frac{\lambda}{\sqrt{3}}(C Q_{5u+d}^A-3 S Q^V)$&
$\frac{\lambda}{\sqrt{3}}(C Q_{u+5d}^A+3 SQ^V)$\\
&$f_{D+}$ &$-C_D \frac{\lambda}{\sqrt{3}}Q_{5u+d}^A$&
$-C_D \frac{\lambda}{\sqrt{3}}Q_{u+5d}^A$\\
&$f_{D-}$ &$C_D \frac{\lambda}{\sqrt{3}} Q_{5u+d}^A$&
$C_D \frac{\lambda}{\sqrt{3}}Q_{u+5d}^A$\\
\hline
\multirow{8}{1em}{$S_{11}(1640)$\\$^4(8)_{1/2}[70,1^-]_1$}&
$f_{-3}$ &0&	0\\
&$ f_{-1}$ &$\frac{\lambda }{\sqrt{6}}R^-_{u+2d}$&
 $\frac{\lambda }{\sqrt{6}}R^-_{2u+d}$\\
&$f_{+1}$ & $\frac{\lambda }{\sqrt{6}}R^+_{u+2d}$&
 $\frac{\lambda }{\sqrt{6}}R^+_{2u+d}$\\
&$f_{+3}$ &0&	0\\
&$f_{0+}$ &$-\sqrt{ \frac{2}{3}} Q_{u+2d}^A \mathcal{B}(3)$&
$-\sqrt{ \frac{2}{3}} Q_{2u+d}^A \mathcal{B}(3)$\\
&$f_{0-}$ &$-\sqrt{ \frac{2}{3}} Q_{u+2d}^A \mathcal{B}(3)$&
$-\sqrt{ \frac{2}{3}} Q_{2u+d}^A \mathcal{B}(3)$\\
&$f_{D+}$ &$-\sqrt{ \frac{2}{3}} Q_{u+2d}^A \mathcal{B}_D(3)$&
$-\sqrt{ \frac{2}{3}} Q_{2u+d}^A \mathcal{B}_D(3)$\\
&$f_{D-}$ &$-\sqrt{ \frac{2}{3}} Q_{u+2d}^A \mathcal{B}_D(3)$&
$-\sqrt{ \frac{2}{3}} Q_{2u+d}^A \mathcal{B}_D(3)$\\
\hline
\newpage
\hline
\multirow{8}{1em}{$D_{13}(1670)$\\$^4(8)_{3/2}[70,1^-]_1$}&
$f_{-3}$ &$ \frac{ 3 \lambda }{\sqrt{10}}R^-_{u+2d}$&
$ \frac{ 3 \lambda }{\sqrt{10}}R^-_{2u+d}$\\
&$f_{-1}$ &$\frac{ \lambda }{\sqrt{30}}R^-_{u+2d}$&
$\frac{ \lambda }{\sqrt{30}}R^-_{2u+d}$\\
&$f_{+1}$ &$-\frac{ \lambda }{\sqrt{30}}R^+_{u+2d}$&
$-\frac{ \lambda }{\sqrt{30}}R^+_{2u+d}$\\
&$f_{+3}$ &$- \frac{ 3 \lambda }{\sqrt{10}}R^+_{u+2d}$&
$- \frac{ 3 \lambda }{\sqrt{10}}R^+_{2u+d}$\\
&$f_{0+}$ &$C \lambda \sqrt{ \frac{2}{15}} Q_{u+2d}^A$&
$C \lambda \sqrt{ \frac{2}{15}} Q_{2u+d}^A$\\
&$f_{0-}$ &$-C \lambda \sqrt{ \frac{2}{15}} Q_{u+2d}^A$&
$-C \lambda \sqrt{ \frac{2}{15}} Q_{2u+d}^A$\\
&$f_{D+}$ &$C_D \lambda \sqrt{ \frac{2}{15}} Q_{u+2d}^A$&
$C_D \lambda \sqrt{ \frac{2}{15}} Q_{2u+d}^A$\\
&$f_{D-}$ &$-C_D \lambda \sqrt{ \frac{2}{15}} Q_{u+2d}^A$&
$-C_D \lambda \sqrt{ \frac{2}{15}} Q_{2u+d}^A$\\
\hline
\multirow{8}{1em}{$D_{15}(1680)$\\$^4(8)_{5/2}[70,1^-]_1$}&
$f_{-3}$ &$-\lambda \sqrt{ \frac{3}{5}} R^-_{u+2d}$&
$-\lambda \sqrt{ \frac{3}{5}}R^-_{2u+d}$\\
&$f_{-1}$ &$-\lambda \sqrt{ \frac{3}{10}} R^-_{u+2d}$&
$-\lambda \sqrt{ \frac{3}{10}} R^-_{2u+d}$\\
&$f_{+1}$ &$-\lambda \sqrt{ \frac{3}{10}}R^+_{u+2d}$&
$-\lambda \sqrt{ \frac{3}{10}} R^+_{2u+d}$\\
&$f_{+3}$ &$-\lambda \sqrt{ \frac{3}{5}} R^+_{u+2d}$&
$-\lambda \sqrt{ \frac{3}{5}} R^+_{2u+d}$\\
&$f_{0+}$ &$-C \lambda \sqrt{ \frac{6}{5}} Q_{u+2d}^A$&
$-C \lambda \sqrt{ \frac{6}{5}}  Q_{2u+d}^A$\\
&$f_{0-}$ &$-C \lambda \sqrt{ \frac{6}{5}} Q_{u+2d}^A$&
$-C \lambda \sqrt{ \frac{6}{5}} Q_{2u+d}^A$\\
&$f_{D+}$ &$-C_D \lambda \sqrt{ \frac{6}{5}} Q_{u+2d}^A$&
$-C_D \lambda \sqrt{\frac{6}{5}} Q_{2u+d}^A$\\
&$f_{D-}$ &$-C_D \lambda \sqrt{ \frac{6}{5}} Q_{u+2d}^A$&
$-C_D \lambda \sqrt{ \frac{6}{5}} Q_{2u+d}^A$\\
\hline
\multirow{8}{1em}{$S_{31}(1620)$\\$^2(10)_{1/2}[70,1^-]_1$}&
$f_{-3}$ &0&	0\\
&$f_{-1}$ &$-\frac{1}{2\sqrt{3}}(6T^- - \lambda\sqrt{2}R^-)$ &
$-\frac{1}{2\sqrt{3}}(6T^- -\lambda\sqrt{2}R^-)$\\
&$f_{+1}$ &$-\frac{1}{2\sqrt{3}}(6T^+-\lambda\sqrt{2}R^+)$ & 
$-\frac{1}{2\sqrt{3}}(6T^+-\lambda\sqrt{2}R^+)$\\
&$f_{+3}$ &0&	0\\
&$f_{0+}$ &$\frac{1}{\sqrt{6}}(Q^A\mathcal{B}(3) + 3 \lambda S Q^V)$&	
$\frac{1}{\sqrt{6}}(Q^A\mathcal{B}(3) + 3 \lambda S Q^V)$\\
&$f_{0-}$ &$\frac{1}{\sqrt{6}}(Q^A\mathcal{B}(3) - 3 \lambda S Q^V)$&
$\frac{1}{\sqrt{6}}(Q^A\mathcal{B}(3) - 3 \lambda S Q^V)$\\
&$f_{D+}$ &$ \frac{1}{\sqrt{6}}  Q^A \mathcal{B}_D(3)$&
$ \frac{1}{\sqrt{6}} Q^A  \mathcal{B}_D(3)$\\
&$f_{D-}$ &$ \frac{1}{\sqrt{6}}  Q^A  \mathcal{B}_D(3)$&
$ \frac{1}{\sqrt{6}} Q^A \mathcal{B}_D(3)$\\
\hline
\multirow{8}{1em}{$D_{33}(1730)$\\$^2(10)_{3/2}[70,1^-]_1$}&
$f_{-3}$ &$-\frac{3}{\sqrt{2}}T^-$&
$-\frac{3}{\sqrt{2}}T^-$\\
&$f_{-1}$ &$-\sqrt{\frac{3}{2}}T^--\frac{\lambda}{\sqrt{3}}R^-$&
$-\sqrt{\frac{3}{2}}T^- -\frac{\lambda}{\sqrt{3}}R^-$\\
&$f_{+1}$ &$\sqrt{\frac{3}{2}}T^+ +\frac{\lambda}{\sqrt{3}}R^+$&
$\sqrt{\frac{3}{2}}T^+ +\frac{\lambda}{\sqrt{3}}R^+$\\
&$f_{+3}$ &$\frac{3}{\sqrt{2}}T^+$&
$\frac{3}{\sqrt{2}}T^+$\\
&$f_{0+}$ &$\frac{\lambda}{\sqrt{3}}(C Q^A-3 S Q^V)$&
$\frac{\lambda}{\sqrt{3}}(C  Q^A - 3 S Q^V)$\\
&$f_{0-}$ &$-\frac{\lambda}{\sqrt{3}}(C Q^A + 3 S Q^V)$&
$-\frac{\lambda}{\sqrt{3}}(C Q^A + 3 S Q^V)$\\
&$f_{D+}$ &$C_D \frac{\lambda}{\sqrt{3}}  Q^A$&
$C_D \frac{\lambda}{\sqrt{3}} Q^A$\\
&$f_{D-}$ &$-C_D \frac{\lambda}{\sqrt{3}} Q^A$&
$-C_D \frac{\lambda}{\sqrt{3}} Q^A$\\
\hline
\newpage
\hline
\multirow{8}{1em}{$P_{11}(1450)$\\$^2(8)_{1/2}[56,0^+]_2$}&
$f_{-3}$ &0&	0\\
&$f_{-1}$ &$\frac{\lambda^2}{2 \sqrt{3}}R^-_{4u-d}$&
$-\frac{\lambda^2}{2 \sqrt{3}}R^-_{u-4d}$\\
&$f_{+1}$ &$-\frac{\lambda^2}{2 \sqrt{3}} R^+_{4u-d}$&
$\frac{\lambda^2}{2 \sqrt{3}} R^+_{u-4d}$\\
&$f_{+3}$ &0&	0\\
&$f_{0+}$ &$\frac{\lambda}{2 \sqrt{3}}(Q_{4u-d}^A\mathcal{B}(2)-3 \lambda S Q_{2u+d}^V)$&
$-\frac{\lambda}{2 \sqrt{3}}(Q_{u-4d}^A\mathcal{B}(2) + 3 \lambda S Q_{u+2d}^V)$\\
&$f_{0-}$ &$-\frac{\lambda}{2\sqrt{3}}(Q_{4u-d}^A\mathcal{B}(2) + 3 \lambda S Q_{2u+d}^V)$&
$\frac{\lambda}{2\sqrt{3}}(Q_{u-4d}^A\mathcal{B}(2) -3 \lambda S Q_{u+2d}^V)$\\
&$f_{D+}$ &$\frac{\lambda}{2 \sqrt{3}}Q_{4u-d}^A \mathcal{B}_D(2)$&
$-\frac{\lambda}{2\sqrt{3}}Q_{u-4d}^A \mathcal{B}_D(2)$\\
&$f_{D-}$ &$-\frac{\lambda}{2\sqrt{3}}Q_{4u-d}^A \mathcal{B}_D(2)$&
$\frac{\lambda}{2 \sqrt{3}}Q_{u-4d}^A \mathcal{B}_D(2)$\\
\hline
\multirow{8}{1em}{$P_{33}(1640)$\\$^4(10)_{3/2}[56,0^+]_2$}&
$f_{-3}$ &$-\frac{\lambda^2}{\sqrt{2}} R^-$&
$-\frac{\lambda^2}{\sqrt{2}}R^-$\\
&$f_{-1}$ &$-\frac{\lambda^2}{\sqrt{6}}R^-$&
$-\frac{\lambda^2}{\sqrt{6}}R^-$\\
&$f_{+1}$ &$-\frac{\lambda^2}{\sqrt{6}}R^+$&
$-\frac{\lambda^2}{\sqrt{6}}R^+$\\
&$f_{+3}$ &$-\frac{\lambda^2}{\sqrt{2}}R^+$&
$-\frac{\lambda^2}{\sqrt{2}}R^+$\\
&$f_{0+}$ &$\lambda \sqrt{ \frac{2}{3}} Q^A\mathcal{B}(2)$&
$\lambda \sqrt{ \frac{2}{3}} Q^A\mathcal{B}(2)$\\
&$f_{0-}$ &$\lambda \sqrt{ \frac{2}{3}} Q^A\mathcal{B}(2)$&
$\lambda \sqrt{ \frac{2}{3}} Q^A\mathcal{B}(2)$\\
&$f_{D+}$ &$\lambda \sqrt{ \frac{2}{3}} Q^A\mathcal{B}_D(2)$&
$\lambda \sqrt{ \frac{2}{3}}Q^A \mathcal{B}_D(2)$\\
&$f_{D-}$ &$\lambda \sqrt{ \frac{2}{3}} Q^A \mathcal{B}_D(2)$&
$\lambda \sqrt{ \frac{2}{3}}Q^A \mathcal{B}_D(2)$\\
\hline
\multirow{8}{1em}{$P_{13}(1740)$\\$^2(8)_{3/2}[56,2^+]_2$}&
$f_{-3}$ &$-\frac{3\lambda}{\sqrt{10}} T^-_{2u+d}$&	
$-\frac{3\lambda}{\sqrt{10}} T^-_{u+2d}$\\
&$f_{-1}$ &$\frac{\lambda}{2\sqrt{15}}(9\sqrt{2}T^-_{2u+d}+2\lambda R^-_{4u-d})$&
$\frac{\lambda}{2\sqrt{15}}(9\sqrt{2}T^-_{u+2d}-2\lambda R^-_{u-4d})$\\
&$f_{+1}$ &$\frac{\lambda}{2\sqrt{15}}(9\sqrt{2}T^+_{2u+d}+2\lambda R^+_{4u-d})$&
$\frac{\lambda}{2\sqrt{15}}(9\sqrt{2}T^+_{u+2d}-2\lambda R^+_{u-4d})$\\
&$f_{+3}$ &$-\frac{3\lambda}{\sqrt{10}}T^+_{2u+d}$&
$-\frac{3\lambda}{\sqrt{10}} T^+_{u+2d}$\\
&$f_{0+}$ &$\frac{\lambda}{\sqrt{15}} (Q_{4u-d}^A\mathcal{B}(5)-3 \lambda S Q_{2u+d}^V)$&
$-\frac{\lambda}{\sqrt{15}} (Q_{u-4d}^A\mathcal{B}(5)+3 \lambda S Q_{u+2d}^V)$\\
&$f_{0-}$ &$\frac{\lambda}{\sqrt{15}} (Q_{4u-d}^A\mathcal{B}(5)+3 \lambda S Q_{2u+d}^V)$&
$-\frac{\lambda}{\sqrt{15}} (Q_{u-4d}^A\mathcal{B}(5)-3 \lambda S Q_{u+2d}^V)$\\
&$f_{D+}$ &$\frac{\lambda}{\sqrt{15}} Q_{4u-d}^A \mathcal{B}_D(5)$&
$-\frac{\lambda}{\sqrt{15}} Q_{u-4d}^A \mathcal{B}_D(5)$\\
&$f_{D-}$ &$\frac{\lambda}{\sqrt{15}} Q_{4u-d}^A \mathcal{B}_D(5)$&
$-\frac{\lambda}{\sqrt{15}} Q_{u-4d}^A \mathcal{B}_D(5)$\\
\hline
\multirow{8}{1em}{$F_{15}(1680)$\\$^2(8)_{5/2}[56,2^+]_2$}&
$f_{-3}$ &$3 \lambda \sqrt{ \frac{2}{5}}T^-_{2u+d}$&
$3 \lambda \sqrt{ \frac{2}{5}} T^-_{u+2d}$\\
&$f_{-1}$ &$ \frac{\lambda}{2\sqrt{5}} (6T^-_{2u+d} - \lambda \sqrt{2}R^-_{4u-d})$&
$ \frac{\lambda}{2\sqrt{5}} (6T^-_{u+2d} + \lambda \sqrt{2} R^-_{u-4d})$\\
&$f_{+1}$ &$-\frac{\lambda}{2\sqrt{5}} (6T^+_{2u+d} - \lambda \sqrt{2}R^+_{4u-d})$&
$ -\frac{\lambda}{2\sqrt{5}} (6T^+_{u+2d} + \lambda \sqrt{2}R^+_{u-4d})$\\
&$f_{+3}$ &$-3 \lambda \sqrt{ \frac{2}{5}}T^+_{2u+d}$&
$-3 \lambda \sqrt{ \frac{2}{5}}T^+_{u+2d}$\\
&$f_{0+}$ &$\frac{\lambda^2}{\sqrt{10}}  (C Q_{4u-d}^A+3 S Q_{2u+d}^V)$&
$-\frac{\lambda^2}{\sqrt{10}}  (C Q_{u-4d}^A-3 S Q_{u+2d}^V)$\\
&$f_{0-}$ &$-\frac{\lambda^2}{\sqrt{10}}  (C Q_{4u-d}^A-3 S Q_{2u+d}^V)$&
$\frac{\lambda^2}{\sqrt{10}}  (C Q_{u-4d}^A+3 S Q_{u+2d}^V)$\\
&$f_{D+}$ &$C_D \frac{ \lambda^2 }{\sqrt{10}}Q_{4u-d}^A$&
$-C_D \frac{ \lambda^2 }{\sqrt{10}}Q_{u-4d}^A$\\
&$f_{D-}$ &$-C_D \frac{ \lambda^2 }{\sqrt{10}}Q_{4u-d}^A$&
$C_D \frac{ \lambda^2 }{\sqrt{10}}Q_{u-4d}^A$\\
\hline
\newpage
\hline
\multirow{8}{1em}{$P_{31}(1920)$\\$^4(10)_{1/2}[56,2^+]_2$}&
$f_{-3}$ &0&	0\\
&$f_{-1}$ &$\frac{\lambda^2}{\sqrt{15}}R^-$&
$\frac{\lambda^2}{\sqrt{15}}R^-$\\
&$f_{+1}$ &$-\frac{\lambda^2}{\sqrt{15}}R^+$&
$-\frac{\lambda^2}{\sqrt{15}}R^+$\\
&$f_{+3}$ &0&	0\\
&$f_{0+}$ &$-\frac{2\lambda}{\sqrt{15}}Q^A \mathcal{B}(5)$&
$-\frac{2\lambda}{\sqrt{15}}Q^A \mathcal{B}(5)$\\
&$f_{0-}$ &$\frac{2\lambda}{\sqrt{15}}Q^A \mathcal{B}(5)$&
$\frac{2\lambda}{\sqrt{15}}Q^A\mathcal{B}(5)$\\
&$f_{D+}$ &$-\frac{2\lambda}{\sqrt{15}}Q^A \mathcal{B}_D(5)$&
$-\frac{2\lambda}{\sqrt{15}}Q^A\mathcal{B}_D(5))$\\
&$f_{D-}$ &$\frac{2\lambda}{\sqrt{15}}Q^A \mathcal{B}_D(5)$&
$\frac{2\lambda}{\sqrt{15}}Q^A \mathcal{B_D}(5)$\\
\hline
\multirow{8}{1em}{$P_{33}(1960)$\\$^4(10)_{3/2}[56,2^+]_2$}&
$f_{-3}$ &$\frac{\lambda^2}{\sqrt{5}}R^-$&
$\frac{\lambda^2}{\sqrt{5}}R^-$\\
&$f_{-1}$ &$-\frac{\lambda^2}{\sqrt{15}}R^-$&
$-\frac{\lambda^2}{\sqrt{15}}R^-$\\
&$f_{+1}$ &$-\frac{\lambda^2}{\sqrt{15}}R^+$&
$-\frac{\lambda^2}{\sqrt{15}}R^+$\\
&$f_{+3}$ &$\frac{\lambda^2}{\sqrt{5}}R^+$&
$\frac{\lambda^2}{\sqrt{5}}R^+$\\
&$f_{0+}$ &$\frac{2\lambda}{\sqrt{15}}Q^A\mathcal{B}(5)$&
$\frac{2\lambda}{\sqrt{15}}Q^A\mathcal{B}(5)$\\
&$f_{0-}$ &$\frac{2\lambda}{\sqrt{15}}Q^A\mathcal{B}(5)$&
$\frac{2\lambda}{\sqrt{15}}Q^A\mathcal{B}(5)$\\
&$f_{D+}$ &$\frac{2\lambda}{\sqrt{15}}Q^A\mathcal{B}_D(5)$&
$\frac{2\lambda}{\sqrt{15}}Q^A\mathcal{B}_D(5)$\\
&$f_{D-}$ &$\frac{2\lambda}{\sqrt{15}}Q^A\mathcal{B}_D(5)$&
$\frac{2\lambda}{\sqrt{15}}Q^A\mathcal{B}_D(5)$\\
\hline
\multirow{8}{1em}{$F_{35}(1920)$\\$^4(10)_{5/2}[56,2^+]_2$}&
$f_{-3}$ &$-3 \lambda^2 \sqrt{ \frac{2}{35}}R^-$&	
$-3 \lambda^2 \sqrt{ \frac{2}{35}}R^-$\\
&$f_{-1}$ &$-\frac{\lambda^2}{\sqrt{35}}R^-$&
$-\frac{\lambda^2}{\sqrt{35}}R^-$\\
&$f_{+1}$ &$\frac{\lambda^2}{\sqrt{35}}R^+$&
$\frac{\lambda^2}{\sqrt{35}}R^+$\\
&$f_{+3}$ &$3 \lambda^2 \sqrt{ \frac{2}{35}}R^+$&	
$3 \lambda^2 \sqrt{ \frac{2}{35}}R^+$\\
&$f_{0+}$ &$-2 C \frac{\lambda^2}{\sqrt{35}}Q^A$&
$-2 C \frac{\lambda^2}{\sqrt{35}}Q^A$\\
&$f_{0-}$ &$2 C \frac{\lambda^2}{\sqrt{35}}Q^A$&
$2 C \frac{\lambda^2}{\sqrt{35}}Q^A$\\
&$f_{D+}$ &$-2 C_D \frac{\lambda^2}{\sqrt{35}}Q^A$&
$-2 C_D \frac{\lambda^2}{\sqrt{35}}Q^A$\\
&$f_{D-}$ &$2 C_D \frac{\lambda^2}{\sqrt{35}}Q^A$&
$2 C_D \frac{\lambda^2}{\sqrt{35}}Q^A$\\
\hline
\multirow{8}{1em}{$F_{37}(1950)$\\$^4(10)_{7/2}[56,2^+]_2$}&
$f_{-3}$ &$\lambda^2 \sqrt{ \frac{2}{7}}R^-$&
$\lambda^2 \sqrt{ \frac{2}{7}}R^-$\\
&$f_{-1}$ &$\lambda^2 \sqrt{ \frac{6}{35}}R^-$&
$\lambda^2 \sqrt{ \frac{6}{35}}R^-$\\
&$f_{+1}$ &$\lambda^2 \sqrt{ \frac{6}{35}}R^+$&
$\lambda^2 \sqrt{ \frac{6}{35}}R^+$\\
&$f_{+3}$ &$\lambda^2 \sqrt{ \frac{2}{7}}R^+$&
$\lambda^2 \sqrt{ \frac{2}{7}}R^+$\\
&$f_{0+}$ &$2 C \lambda^2 \sqrt{ \frac{6}{35}} Q^A$&
$2 C \lambda^2 \sqrt{ \frac{6}{35}} Q^A$\\
&$f_{0-}$ &$2 C \lambda^2 \sqrt{ \frac{6}{35}}Q^A$&
$2 C \lambda^2 \sqrt{ \frac{6}{35}} Q^A$\\
&$f_{D+}$ &$2 C_D \lambda^2 \sqrt{ \frac{6}{35}} Q^A$&
$2 C_D \lambda^2 \sqrt{ \frac{6}{35}} Q^A$\\
&$f_{D-}$ &$2 C_D \lambda^2 \sqrt{ \frac{6}{35}} Q^A$&
$2 C_D \lambda^2 \sqrt{ \frac{6}{35}} Q^A$\\
\hline
\newpage
\hline
\multirow{8}{1em}{$P_{11}(1710)$\\$^2(8)_{1/2}[70,0^+]_2$}&
$f_{-3}$ &0&	0\\
&$f_{-1}$ &$-\frac{\lambda^2}{2 \sqrt{6}}R^-_{5u+d}$&
$-\frac{\lambda^2}{2 \sqrt{6}}R^-_{u+5d}$\\
&$f_{+1}$ &$\frac{\lambda^2}{2 \sqrt{6}}R^+_{5u+d}$&
$\frac{\lambda^2}{2 \sqrt{6}}R^+_{u+5d}$\\
&$f_{+3}$ &0&	0\\
&$f_{0+}$ &$-\frac{\lambda}{2 \sqrt{6}}(Q_{5u+d}^A\mathcal{B}(2)-3 \lambda SQ^V)$&
$-\frac{\lambda}{2 \sqrt{6}}(Q_{u+5d}^A\mathcal{B}(2) + 3 \lambda S Q^V)$\\
&$f_{0-}$ &$\frac{\lambda}{2 \sqrt{6}}(Q_{5u+d}^A\mathcal{B}(2)+ 3 \lambda S Q^V)$&
$\frac{\lambda}{2 \sqrt{6}}(Q_{u+5d}^A\mathcal{B}(2) - 3 \lambda S Q^V)$\\
&$f_{D+}$ &$-\frac{\lambda}{2 \sqrt{6}}Q_{5u+d}^A\mathcal{B}_D(2)$&
$-\frac{\lambda}{2 \sqrt{6}}Q_{u+5d}^A \mathcal{B}_D(2)$\\
&$f_{D-}$ &$\frac{\lambda}{2 \sqrt{6}}Q_{5u+d}^A \mathcal{B}_D(2)$&
$\frac{\lambda}{2 \sqrt{6}}Q_{u+5d}^A \mathcal{B}_D(2)$\\
\hline
\multirow{8}{1em}{$F_{17}(1970)$\\$^4(8)_{7/2}[70,2^+]_2$}&
$f_{-3}$ &$-\frac{\lambda^2}{\sqrt{14}}R^-_{u+2d}$&
$-\frac{\lambda^2}{\sqrt{14}}R^-_{2u+d}$\\
&$f_{-1}$ &$-\lambda^2 \sqrt{ \frac{3}{70}}R^-_{u+2d}$&
$-\lambda^2 \sqrt{ \frac{3}{70}}R^-_{2u+d}$\\
&$f_{+1}$ &$-\lambda^2 \sqrt{ \frac{3}{70}}R^+_{u+2d}$&
$-\lambda^2 \sqrt{ \frac{3}{70}}R^+_{2u+d}$\\
&$f_{+3}$ &$-\frac{\lambda^2}{\sqrt{14}}R^+_{u+2d}$&
$-\frac{\lambda^2}{\sqrt{14}}R^+_{2u+d}$\\
&$f_{0+}$ &$-C \lambda^2 \sqrt{ \frac{6}{35}} Q_{u+2d}^A$&
$-C \lambda^2 \sqrt{ \frac{6}{35}}Q_{2u+d}^A$\\
&$f_{0-}$ &$-C \lambda^2 \sqrt{ \frac{6}{35}}Q_{u+2d}^A$&
$-C \lambda^2 \sqrt{ \frac{6}{35}}Q_{2u+d}^A$\\
&$f_{D+}$ &$-C_D \lambda^2 \sqrt{ \frac{6}{35}} Q_{u+2d}^A$&
$-C_D \lambda^2 \sqrt{ \frac{6}{35}} Q_{2u+d}^A$\\
&$f_{D-}$ &$-C_D \lambda^2 \sqrt{ \frac{6}{35}}Q_{u+2d}^A$&
$-C_D \lambda^2 \sqrt{ \frac{6}{35}} Q_{2u+d}^A$\\
\hline
\caption{Helicity amplitudes for all baryonic resonances included in \lstinline{GENIE}.\label{long}}
\end{longtable}
}
\label{sec:Helicity_table}

\bibliographystyle{JHEP}
\bibliography{resonant-bdm}
\end{document}